\documentclass[aps,reprint,prd,superscriptaddress,floatfix,nofootinbib,longbibliography]{revtex4-2}
\usepackage{graphicx,longtable,mathrsfs,color,array}
\usepackage{array, makecell}
\usepackage[usenames,dvipsnames]{xcolor}
\usepackage{amssymb,amsmath,mathtools,mathrsfs,slashed}
\usepackage{quantikz}
\usepackage{float}
\usepackage{placeins}
\usepackage{dsfont}
\usepackage[dvipsnames]{xcolor}
\usepackage[pdftex,
            pdftitle={String Breaking in Z2 LGT},
            pdfauthor={Constantia Alexandrou, Andreas Athenodorou, Kostas Blekos, Georgios Polykratis, Stefan Kuehn},
            bookmarks,
            colorlinks,
            linkcolor=myblue,
            citecolor=mymagenta,
            menucolor=black,
            urlcolor=myblue,
            plainpages=false,
            pdfpagelabels,
            hypertexnames=false]{hyperref}
\usepackage{orcidlink}
\usepackage{comment}
\definecolor{mymagenta}{RGB}{200, 0, 100}
\definecolor{myblue}{RGB}{45, 48, 146}
\definecolor{mygreen}{RGB}{0, 126, 0}
\definecolor{myorange}{RGB}{255, 136, 19}

\newcommand{\sm}{\ensuremath{\sigma^-}}

\newcommand{\Zt}{\ensuremath{\mathds{Z}_2}}
\renewcommand\vec{\boldsymbol}

\begin{document}
\title{Realizing string breaking dynamics in a $\Zt$ lattice gauge theory on quantum hardware}

\author{Constantia Alexandrou\orcidlink{0000-0001-9136-3621}}
\affiliation{Computation-based Science and Technology Research Center, The Cyprus Institute, Cyprus}

\author{Andreas Athenodorou\orcidlink{0000-0003-4600-4245}}
\affiliation{Computation-based Science and Technology Research Center, The Cyprus Institute, Cyprus}

\author{Kostas Blekos\orcidlink{0000-0002-6777-2107}}
\affiliation{Computation-based Science and Technology Research Center, The Cyprus Institute, Cyprus}

\author{Georgios Polykratis}
\affiliation{Computation-based Science and Technology Research Center, The Cyprus Institute, Cyprus}

\author{Stefan Kühn\orcidlink{0000-0001-7693-350X}}
\affiliation{Deutsches Elektronen-Synchrotron DESY, Platanenallee 6, 15738 Zeuthen, Germany}

\date{\today}

\begin{abstract}
    We investigate static and dynamical aspects of string breaking in a $\Zt$ lattice gauge theory coupled to Kogut-Susskind staggered fermions. Using Tensor Network simulations, we demonstrate that the static potential as well as the site-resolved configuration of the matter sites and gauge links allows us to identify the regimes in which string breaking occurs. Furthermore, we develop a variational quantum eigensolver that allows for reliably preparing the ground state of the theory in both the absence and presence of static charges and to capture the static aspects of the phenomenon. Carrying out state preparation on real quantum hardware for up to 19 qubits, we demonstrate its suitability for current quantum devices. In addition, we study the real-time dynamics of a flux tube between two static charges using both Tensor Networks and quantum hardware. Using a trotterization for the time-evolution operator, we are able to show that the breaking process starts with the creation of charges inside the string. These eventually redistribute towards the static charges and screen them, which leads to the breaking of the flux tube.
\end{abstract}

\maketitle

\section{Introduction\label{sec:introduction}}
String breaking is one of the most fundamental phenomena in confining gauge field theories. In the Standard Model, it is an effect described by Quantum Chromodynamics (QCD), where a quark and an antiquark joined together by a flux-tube will experience a linearly rising potential with increasing distance. From a certain separation on, it becomes energetically favorable to break the flux-tube by forming two mesons - two separate flux-tubes joining pairs of quark-antiquark. This occurs when the energy of the flux-tube equals twice the mass of the lightest meson, allowing the vacuum to dynamically create pairs of quarks and antiquarks. 

Lattice gauge theories~\cite{Wilson1974,Kogut1975} offer a robust framework that allows physicists to investigate phenomena related to quark confinement, such as the formation and breaking of strings. In LGTs, spacetime is discretized, with fermionic degrees of freedom placed at lattice sites and gauge fields assigned to the links joining lattice sites. Since LGTs are formulated on a discrete spacetime, they provide an ideal setting for numerical calculations of the equilibrium properties of gauge fields. Conventionally, such numerical calculations are performed using Monte Carlo simulations in Euclidean spacetime.  These are highly successful for studying the static potential and string breaking~\cite{Bali1992,Philipsen1998,Bali2005,Pepe2009}, mass spectra~\cite{Duerr2008} phase diagrams and other static properties. String breaking is observed in lattice gauge theories (LGTs), when dynamical quarks are being considered~\cite{Bali:2005fu,Bulava:2019iut}.However, Monte Carlo simulations face a major challenge when the action cannot be rendered real resulting in a sign problem, which hinders an efficient Monte Carlo sampling in certain parameter regimes, such as when the chemical potential is non-zero. Hence, there is an ongoing interest in alternative methods overcoming this limitations such as Tensor Networks (TNs)~\cite{Banuls2018a,Banuls2020,Magnifico2024}, or equivariant learning networks~\cite{Luo2021,Favoni2022}. 

Quantum computing is widely regarded as the next evolution of computational infrastructure allowing the problems intractable by conventional computers to be studied. In particular, it provides a promising avenue towards simulating LGTs in regimes that are not accessible with conventional MC methods, such as out-of-equilibrium dynamics or regimes where the sign problem occurs~\cite{Banuls2020,DiMeglio2023,Bauer2023a,Bauer2024}. However, practical implementation of LGTs will require adapting the immense complexity of the lattice formulation to quantum hardware. While it may be too early to discuss realistic simulations of gauge theories relevant for the Standard Model, first successful demonstrations have been carried out using toy models that preserve similar gauge symmetries and can be formulated in fewer dimensions~\cite{Kokail2018,Atas2022,Farrell2023,Chai2023a,Angelides2023a,Farrell2024}. In this work, we explore aspects of string breaking, which can be observed in a quantum setup by studying a simpler gauge model, namely the $\Zt$ model in 1+1 dimensions.

Recent advancements in simulating string breaking and confinement on quantum computers have been reported in several studies, each contributing valuable insights into these phenomena. For instance, Ref.~\cite{Surace2024} explores string-breaking dynamics in Ising-type models by examining a system of (1+1)-dimensional quantum spin chains. The authors induce string breaking through both adiabatic and controlled diabatic processes, providing a deeper understanding of real-time dynamics. Their findings suggest that the process can be effectively modeled as a transition between two states in a quantum system, analogous to the Landau-Zener process.

In a related study, Ref.~\cite{De:2024smi} delves into string-breaking phenomena within a (1+1)-dimensional $\Zt$ lattice gauge theory using a fully programmable trapped-ion quantum simulator. The authors investigate how confinement influences isolated charges, showing that, in the absence of string tension, the charges disperse freely. However, as string tension increases, they exhibit localized, coherent oscillations. The study further explores the dynamics of a string stretched between two static charges and, following a sudden increase in string tension, observes the formation of charge pairs near the endpoints of the string, which then propagate into the bulk. This highlights the potential for quantum simulators to explore both confinement and string-breaking processes.

Similarly, Ref.~\cite{Mildenberger2025} uses a superconducting quantum processor to investigate confinement in a (1+1)-dimensional $\Zt$ lattice gauge theory. The authors employ a three-qubit gate to simulate long-term dynamics, demonstrating charge confinement and string formation via gauge constraints. Notably, while this study reveals important aspects of gauge symmetry's influence on system behavior, including the freezing of dynamics in a U(1) setup, it does not observe string breaking, indicating a gap in the ability to simulate these complex phenomena on current quantum hardware.

Further extending the scope to non-Abelian gauge theories, Ref.~\cite{Ciavarella2024} examines string breaking in the heavy quark limit of a (1+1)-dimensional SU(2) lattice gauge theory. The study presents a method for simulating non-Abelian gauge theories by truncating the Hilbert space and developing scalable variational circuits for vacuum and meson state preparation. Using $104$ qubits on IBM's Heron quantum computer, the authors successfully simulate meson production during time evolution and observe inelastic pair production. This work highlights the feasibility of simulating non-Abelian gauge theories on quantum computers, with a focus on string breaking and quark-antiquark dynamics.

Ref.~\cite{Charles2024} tackles the challenge of simulating (1+1)-dimensional $\Zt$ lattice gauge theories on quantum computers while addressing hardware noise. Through advanced error mitigation techniques-such as readout error mitigation, randomized compiling, rescaling, and dynamical decoupling-the authors improve the reliability of their simulations, extending the accurate simulation time by a factor of six. This study underscores the critical role of error mitigation in enabling quantum simulations of lattice gauge theories.

Finally, the Google Quantum AI research team, in collaboration with other researchers, presents Ref.~\cite{Cochran2024}, which simulates (2+1)-dimensional $\Zt$ lattice gauge theory on a quantum processor. This work investigates the dynamics of charges and strings, focusing on the transition from deconfined to confined dynamics induced by an external magnetic field. The study identifies two distinct regimes of confinement: weak confinement with strong transverse fluctuations and strong confinement, where these fluctuations freeze. Importantly, this research presents a new approach for simulating particle and string dynamics, including string breaking phenomena, on a quantum processor.

In this work, we focus on simulating string breaking in a (1+1)-dimensional $\Zt$ LGT~\cite{Horn1979} coupled to Kogut-Susskind  staggered fermions~\cite{Kogut1975} on a digital quantum computer. Starting from the Hamiltonian formulation, we investigate both the static and dynamical properties of the system. In particular, we develop a gauge-invariance preserving variational quantum eigensolver (VQE), which allows us to prepare the ground state of the theory and to identify the parameter regimes in which string breaking occurs. In a second step, we simulate string dynamics to explore the features of string breaking in real time. We use TNs as a testbed for examining various parameter regimes, before carrying out simulations on IBM's quantum hardware with up to 19 qubits. Moreover, we discuss a path forward to scaling up to larger system sizes.

The rest of the article is structured as follows: In Sec.~\ref{sec:model_and_methods}, we introduce the $\Zt$ LGT  and the numerical methods that we used to study  properties associated to string breaking. Section~\ref{sec:numerical_results} contains our numerical results and Sec.~\ref{sec:conclusions} contains our conclusions. 

\section{Model and methods\label{sec:model_and_methods}}

\subsection{$\Zt$ lattice gauge theory}
For our simulations, we adopt the Hamiltonian lattice formulation of $\Zt$ LGTs~\cite{Horn1979} coupled to Kogut-Susskind staggered fermions~\cite{Kogut1975} in 1+1 dimensions. The Hamiltonian for a system with open boundaries and $L$ lattice sites reads
\begin{equation}    
    H = H_\text{kin} + H_m + H_\text{el},         
    \label{eq:Z2_Hamiltonian_fermionic}
\end{equation}
where the individual terms are given by
\begin{align}      
    H_\text{kin} &=J\sum_{l=1}^{L-1}\left(\phi_{l}^\dagger Z_{l,l+1}\phi_{l+1} +\text{h.c.}\right), \label{eq:Z2_Hamiltonian_fermionic_kinetic}\\
    H_m &= m \sum_{l=1}^{L}(-1)^l \phi^\dagger_{l}\phi_{l},\label{eq:Z2_Hamiltonian_fermionic_mass}\\
    H_\text{el} &= \varepsilon \sum_{l=1}^{L-1}X_{l,l+1}.\label{eq:Z2_Hamiltonian_fermionic_electric}        
\end{align}
In the Eqs.~\eqref{eq:Z2_Hamiltonian_fermionic_kinetic} - \eqref{eq:Z2_Hamiltonian_fermionic_electric}, $\phi_l$ is a single-component fermionic field describing the matter at site $l$, and $Z_{l,l+1}$, $X_{l,l+1}$ are Pauli matrices acting on the gauge links joining the matter sites $l$ and $l+1$. $H_\text{kin}$ is the kinetic term for the fermions with hopping parameter $J$,
$H_m$ is the staggered mass term with bare fermion mass $m$, and $H_\text{el}$ is the ``electric'' energy of the gauge field\footnote{Note that only in the limit $N\to\infty$ i.e.\ $\mathds{Z}_N \xlongrightarrow{N \to \infty} U(1)$~\cite{Luo:2023cjv} Quantum Electrodynamics is  recovered. Nevertheless, we will refer to the last term in the Hamiltonian as the electric energy.} with parameter $\varepsilon$. Since the staggered formulation essentially corresponds to distributing the  upper (lower) components of the Dirac spinor to even (odd) lattice sites~\cite{Susskind1977}, we will assume $L$ to be even for the rest of the paper.

The physical states $\ket{\psi}$ of the Hamiltonian have to fulfill the Gauss law,$\forall l$ $G_l\ket{\psi} = q_l\ket{\psi}$, where 
\begin{align}
    G_l &= X_{l-1,l} p_lX_{l,l+1} 
    \label{eq:Z2_Gauss_law}
\end{align}
and $p_l =\exp\left(i\pi \mathcal{Q}_l\right)$ with $\mathcal{Q}_l = \phi_{l}^\dagger\phi_{l} - \left(1-(-1)^l\right)/2$ the staggered fermionic charge acting on site $l$, and $q_l=\pm 1$. Since we work with open boundaries, we assume that the gauge field at the boundaries corresponds to $X_{0,1}=X_{L-1,L}=-1$. Note that the modular charge $p_l$ takes the value $1$ if the staggered charge vanishes, and $\exp(\pm i \pi) = -1$ if the staggered charge is nonzero. Thus, we can interpret $q_l = -1$ as a static external charge at site $l$, whereas $q_l=1$ corresponds to the case where no external charge is present. 

For our numerical simulations, it is convenient to translate the fermionic degrees of freedom to spins using a Jordan-Wigner transformation, $\phi_l = \prod_{k<l}\left(iZ_k\right) \sm_l$~\cite{Hamer1997}, where $\sigma^\pm = (X\pm i Y)/2$ and Pauli matrices with a single index act on the corresponding matter site. The spin formulation of the fermionic terms in Eqs.~\eqref{eq:Z2_Hamiltonian_fermionic_kinetic} and \eqref{eq:Z2_Hamiltonian_fermionic_mass} is given by
\begin{align}
    H_\text{kin} &= -\frac{J}{2} \sum_{l=1}^{L-1} \left(X_l Z_{l,l+1}Y_{l+1} - Y_l Z_{l,l+1}X_{l+1}\right),\label{eq:Z2_Hamiltonian_spins_kinetic} \\
    H_m &=\frac{m}{2}\sum_{l=1}^{L} (-1)^l\left(\mathds{1} + Z_{l}\right). \label{eq:Z2_Hamiltonian_spins_mass}    
\end{align}
Correspondingly, the Gauss law in spin formulation can be expressed as 
\begin{align}
    G_l = (-1)^{l+1}X_{l-1,l}Z_lX_{l,l+1},
    \label{eq:Z2_Gauss_law_spins}
\end{align}
where we have used that $p_l$ evaluates to $(-1)^{l+1} Z_l$ after the Jordan-Wigner transformation. Hence, one arrives at a form of the theory that only involves Pauli matrices that can be directly mapped to a qubit-based quantum device or straightforwardly addressed with TN methods.

\subsection{String breaking in the $\Zt$ model}
\begin{figure}[htp!]
     \centering
     \includegraphics[width=\linewidth]{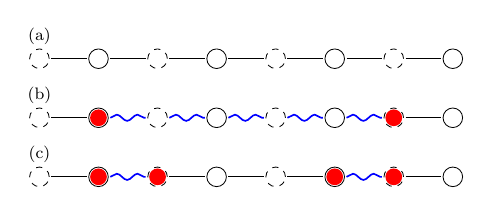}
     \caption{(a) Illustration of the noninteracting ground state, where the dashed (solid) circles denote to odd (even) matter sites, the black horizontal lines to a gauge link in state $\ket{-}$. (b) Flux tube of excited links in the state $\ket{+}$ are indicated by the blue wiggly lines between two charges displayed as red filled circles. (c) Configuration creating two additional charges, which allow for breaking the string and forming two mesons, each corresponding to two charges connected by an excited link.}
     \label{fig:string_breaking_illustration}
\end{figure}

Despite its simplicity, the $\Zt$ model shows string breaking, in analogy to what is observed in more complicated gauge theories, such as Quantum Chromodynamics (QCD). An intuitive picture can be obtained from the non-interacting case, $J\to 0$. In this limit, the kinetic term can be neglected, and the gauge invariant ground state of the Hamiltonian given in Eq.~\eqref{eq:Z2_Hamiltonian_fermionic} in the sector of vanishing static charges, $\forall l$ $q_l=1$, is given by 
\begin{align}
    \begin{aligned}
    \ket{\psi_\text{ni}}
    &= \ket{\bf 1}\ket{-}\ket{\bf 0}\ket{-}\ket{\bf 1}\ket{-}\ket{\bf 0}\dots \ket{-}\ket{\bf 0}\\
    &\equiv \ket{0}\ket{-}\ket{1}\ket{-}\ket{0}\ket{-}\ket{1}\dots \ket{-}\ket{0}.
    \end{aligned}
    \label{eq:sc_state}
\end{align}
In the expression of Eq.~\ref{eq:sc_state}, the numbers in bold face in the first line correspond to the fermionic occupation of the matter site, and $\ket{-}$ refers to the gauge links being in the eigenstate with eigenvalue $-1$ of the Pauli-$X$ operator. The second line gives the equivalent spin representation after translating the fermionic matter to spins, where $\ket{0}$ ($\ket{1}$) is the eigenstate with eigenvalue +1 (-1) of the Pauli-$Z$ matrix. Note that for the state in Eq.~\eqref{eq:sc_state} on finds $\mathcal{Q}_l=0$ for all sites, hence there are no charges present (see also Fig.~\ref{fig:string_breaking_illustration}(a)). 

In order to create two charges, one has to empty (occupy) a previously occupied (empty) site in the ground state given in Eq.~\eqref{eq:sc_state}, or correspondingly flip a pair of spins. In order to maintain gauge invariance, the links between the charges have to be excited, thus forming an electric flux string in the system (c.f.\ Fig.~\ref{fig:string_breaking_illustration}(b)). From a formal point of view, a pair of charges at a distance $d$ starting at site $l$ is created by applying a string operator
\begin{align}
    S_{ld} = \phi_l^\dagger Z_{l,l+1}\dots Z_{l+d-1,l+d}\phi_{l+d}
    \label{eq:string_operator}
\end{align}
or the hermitian conjugate thereof to $\ket{\psi_\text{ni}}$. The resulting state has an excess energy of $2m + 2d\varepsilon$ because of the the two fermionic charges at the end of the string and the excitation of  the links in between from the $\ket{-}$ state to the state $\ket{+}$, which increases the electric energy by two units for each link. Since the excess energy grows linearly with $d$, from a certain length on, it will be energetically favorable to create another pair of charges that allows for shortening the flux tube and breaking it into two mesons (see Fig.~\ref{fig:string_breaking_illustration}(c)). In the noninteracting case, this threshold is reached when an additional contribution of $2m$ plus the two excited links forming the mesons is energetically more favorable than the string. From the simple noninteracting case, we can deduce the critical distance $d_c$ at which we expect the string to break, which is given by
\begin{align}
    d_c = 1 + \frac{m}{\varepsilon}.
    \label{eq:critical_distance_sc}
\end{align}
While the discussion above focuses on the noninteracting case, a similar picture persists in the interacting case~\cite{Buyens2015,Kuehn2015,Sala2018}.

A common way to probe string breaking in a static calculation is to solve the theory in the presence of a pair of static charges. To this end, one computes the ground state for varying distances $d$ of the static charges and monitors the static potential $V(d) = E_d - E_0$, i.e.\ the excess energy of the ground state with static charges $E_d$ at distance $d$ compared to the ground state $E_0$ without static charges. Following the same arguments presented above, one expects that  $V(d)$ increases linearly with $d$ as long as the string is present and becomes independent of $d$ as soon as the string breaks. In our numerical simulations we will compute the static potential and use it as an indicator to identify parameter regimes where string breaking takes place.

In addition, the phenomenon can also be studied fully dynamically by preparing the interacting ground state of the theory and creating a string. Subsequently, this state can be evolved in time, and the evolution of the flux tube can be monitored~\cite{Hebenstreit2013,Buyens2015,Kuehn2015,Ciavarella2024}. While this is not possible with conventional MC methods, TN and quantum computers allow for simulating the evolution of the system.

\subsection{Numerical methods}

To address the model numerically, we use two methods: i) we address it using matrix product states (MPS), a particular kind of one-dimensional TNs~\cite{Schollwoeck2011,Orus2014a}; and ii)  we propose a VQE protocol for static ground state calculations as well as trotterization of the time-evolution operator to simulate the model on a quantum device. In the following, we review the different numerical techniques we utilize for our simulations.

\subsubsection{Matrix product states}
In order to explore the model numerically, also at system sizes and time scales which cannot be reached with current commercially available quantum devices, we use numerical techniques based on TNs. Starting from the wave function of a quantum many-body system with $N$ sites written in the basis of tensor products of local $K$-dimensional bases $\{\ket{i_k}\}_{k=1}^K$,
\begin{align}
    \ket{\psi} = \sum_{i_1,\dots,i_N=1}^K c_{i_1,\dots,i_N}\ket{i_1}\otimes\dots \otimes\ket{i_N},
\end{align}
the basic idea of TNs is to represent the exponentially big tensor $c_{i_1,\dots,i_N}\in\mathds{C}^{K^N}$ as a contraction of a network of smaller pieces~\cite{Orus2014a}. A particular decomposition is to express $c_{i_1,\dots,i_N}$ as a product of complex matrices $A_k^{i_k}$
\begin{align}
    c_{i_1,\dots,i_N} = A_1^{i_1} A_2^{i_2} \dots A_N^{i_N}
    \label{eq:MPS_decomposition}
\end{align}
where the $A_k^{i_k}$ are complex square matrices for $1<k<N$, and $A_1^{i_1}$ ($ A_N^{i_N}$) is a complex row (column) vector. Thus, for a given set of physical indices, the right hand side of Eq.~\eqref{eq:MPS_decomposition} evaluates to a complex number. This decomposition is known as an MPS with open boundary conditions~\cite{Schollwoeck2011,Orus2014a,Silvi2019a}. While the maximum matrix size required for representing a general state scales is $K^{\lfloor N/2\rfloor}$, where $\lfloor N/2\rfloor$ is the largest integer smaller than $N/2$, quantum information theory has shown that low-energy states of many physically relevant Hamiltonians can be efficiently described with matrices whose size does not scale exponentially in $N$. The reason such a description is successful is that these states are generally only moderately entangled, and TNs provide an efficient parametrization of the slightly entangled subspace~\cite{Hastings2007,Schuch2008,Wolf2008,Eisert2010,Brandao2014}.

Fixing the maximum matrix size, also called the bond dimension of the MPS, to  $D$ one obtains a family of states that can be used for numerical computations. These states can be used as variational Ansatz to approximate the ground state of a given Hamiltonian. To this end the tensors $A_i^{i_k}$ are updated iteratively, one at a time, while keeping the others fixed. The optimal tensor in each step is found by determining the ground state of an effective Hamiltonian that describes the interaction between the site to be updated and its environment. Starting from the left boundary and sweeping back and forth through the tensors, until the energy change is smaller than a specified tolerance, the final MPS provides an approximation for the ground state. Subsequently arbitrary expected values of (local) observables such as Pauli operators can be measured. Alternatively, starting from an MPS wave function one can compute the time evolution by applying a trotterized approximation of the time evolution operator. Details on the numerical algorithms can be found in the reviews in Refs.~\cite{Schollwoeck2011,Orus2014a,Silvi2019a}. For our numerical MPS calculations, we use the ITensor library in Julia~\cite{ITensor,ITensor-r0.3}.

Moreover, given an MPS, it is possible to efficiently compute  the reduced density operator $\rho_{s}$ describing a contiguous subsystem of the entire system. This, in turn, allows for computing the von Neumann entropy $S = - \text{tr}(\rho_{s} \log \rho_{s})$, which provides a measure for the quantum correlations between the subsystem and its environment~\cite{Schollwoeck2011,Orus2014a,Silvi2019a}.

\subsubsection{VQE for ground state preparation and real-time evolution on a quantum device\label{subsec:VQE}}
In order to approximate the ground state of the theory on a quantum device and to prepare initial states for studying the evolution of a flux string over time, we use the VQE approach~\cite{Peruzzo2014}.  To this end, we need to design a parametric Ansatz circuit $U(\vec{\theta})$ that prepares the state $\ket{\psi(\vec{\theta})} = U(\vec{\theta}) \ket{\psi_\text{init}}$, where $\vec{\theta}$ is the vector containing the parameters to be optimized throughout the course of the VQE, and $\ket{\psi_\text{init}}$ is an initial state that can be easily realized on the quantum device. The optimal parameters are found by minimizing the expectation value
\begin{equation}
    E(\vec{\theta}) \;=\; \bra{\psi(\vec{\theta})} H \ket{\psi(\vec{\theta})},    
\end{equation}
where the expected value of $H$ is measured on the quantum device and a classical optimizer is used to update $\vec{\theta}$. Repeating this procedure iteratively, until convergence is reached in the change in energy below a desired threshold, the final state represents an approximation for the ground state of $H$. Note that the Hamiltonian of the theory we study after Jordan-Wigner transformation is a sum of Pauli terms, which can be directly measured on a qubit-based quantum device.

\begin{figure*}[htp!]
    \centering
    \includegraphics[width=0.8\textwidth]{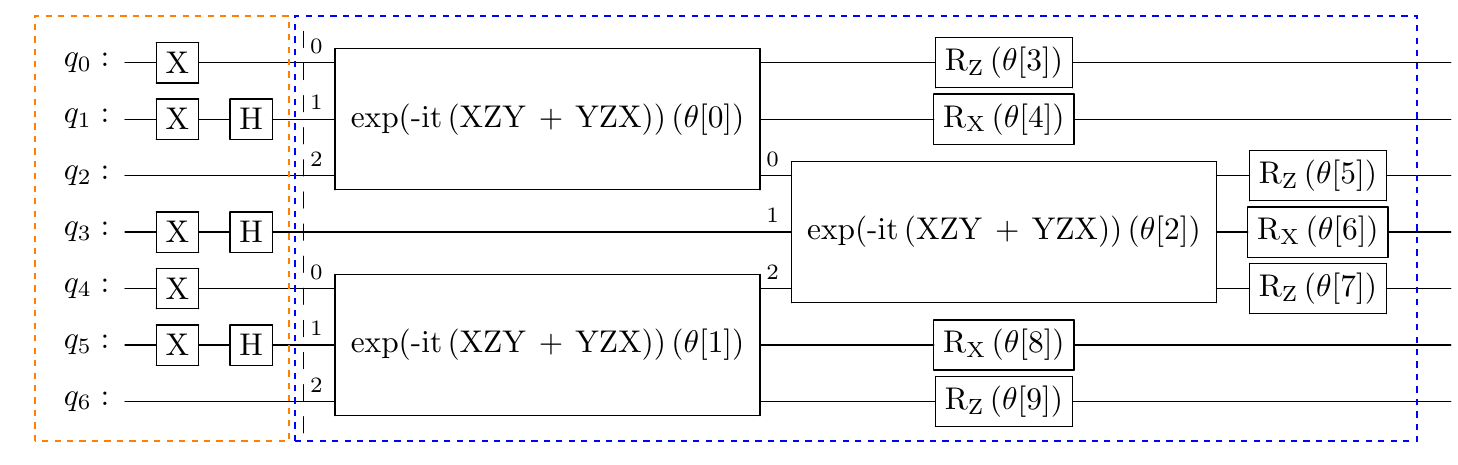}
    \caption{Ansatz circuit used for VQE. The part in the orange box is used to prepare $\ket{\psi_\text{init}}$. The gates inside the blue box correspond to a single layer of the VQE Ansatz, which consists of the trotterized kinetic part with first the terms starting on odd matter sites, followed by the kinetic terms starting on even matter sites and a layer of $R_X$ ($R_Z$) rotation gates on the gauge links (matter sites).}
    \label{fig:circuit_no1stlayer}
\end{figure*}

A particular challenge arises from the fact that one has to ensure the Ansatz circuit $U(\vec{\theta})$ only explores the physically relevant subspace of gauge invariant states fulfilling Eq.~\eqref{eq:Z2_Gauss_law_spins}. While for one-dimensional LGTs with open boundaries the gauge fields can be integrated out, and one can find a formulation directly restricted to this subspace, this leads to a Hamiltonian with long-range interactions~\cite{Sala2018}. Implementing the dynamics under such a Hamiltonian typically results in deep circuits, which would render the dynamical simulation of string breaking later on challenging on current quantum hardware. Alternatively, one could enforce gauge invariance by adding an energy-penalty term to the Hamiltonian that vanishes only for gauge invariant states. However, such an approach is prone to lead to barren plateaus in the VQE~\cite{Cerezo2021,Larocca2024}. Here, we follow the approach outlined in Ref.~\cite{Mazzola2021} and construct a gauge-preserving Ansatz by using individual gauge-invariant Hamiltonian terms as resources to generate unitary operations. Specifically, we construct our circuit from exponentials of each of  the electric, mass and kinetic terms appearing in the Hamiltonian of Eq.~\ref{eq:Z2_Hamiltonian_fermionic}:
\begin{equation}
    e^{-i \alpha\,H_\text{el}}, 
    \quad
    e^{-i \alpha \,H_{m}}, 
    \quad   
    e^{-i \alpha \,H_{\text{kin}}},
\end{equation}
where $\alpha$ is a real parameter. The first two exponentials correspond up to phase factors to a series of standard single-qubit Pauli rotation gates $R_P(\theta)=\exp(-i \theta/2 P)$, with $P$ the Pauli-$X$ ($Z$) matrix for gauge (matter) sites. The third exponential $U_\text{kin}$, in the Trottier approximation, is written into exponentials of non-overlapping three-body Pauli terms that can be decomposed into standard single- and two-qubit gates and implemented on quantum hardware. Note that each of the terms in $H_{\text{kin}}$ is itself gauge invariant, thus the trotterization does not lead to a gauge violating Ansatz. Combining these building blocks into layers and choosing a gauge-invariant state for $\ket{\psi_\text{init}}$, the Ansatz will only explore the physical sector and the gauge symmetry remains intact at every intermediate stage. More specifically, the Ansatz we use is illustrated in Fig.~\ref{fig:circuit_no1stlayer}, where the exponentials of the respective parts of the Hamiltonian are each assigned a separate variational parameter to each gate to enhance expressiveness. 
In practice, circuit depth and parameter count must be balanced against hardware noise, making it useful to exploit spatial symmetries (e.g.\ translation invariance) to reduce redundant parameters in the bulk of the lattice, especially for larger systems (see Appendix~\ref{app:symmetries} for a discussion). 
    
For faster convergence, we initialize $\ket{\psi_\text{init}}$ in a known gauge-invariant reference state. Depending on the regime, convenient choices are the states in the noninteracting limit, $J\to 0$, representing either the ground state without static charges, a flux tube, or two mesons. These states can be easily prepared with only Pauli-$X$ and Hadamard gates (see Fig.~\ref{fig:new_initialization_circuit} in Appendix~\ref{app:circuit_initialization} for an illustration how to realize such states).  

For each VQE iteration, we obtain $E(\boldsymbol{\theta})$ from the individual Pauli terms of $H$, where we typically accumulate $\sim 10^4$ measurements. This number can be tuned based on the noise characteristics of the device. For the classical optimizer updating the parameters based on the measurement outcome we use COBYLA.

\subsubsection{Real-time evolution on a quantum device\label{subsec:qc_evo}}
To study the dynamical aspects of  string breaking with a quantum computer, we first compute the ground state of the system without static charges. Subsequently, we create a string excitation of the ground state by applying the operator $Z_{l,l+1}\dots Z_{l+d-1,l+d}$. This operator corresponds to Eq.~\eqref{eq:string_operator} up to the fermionic parts at the end and excites the links joining sites $l$ to $l + d$. Hence, applying the string operator to the ground state, the Gauss law at the beginning and the end of the string yields $-1$, implying a static external charge at sites $l$ and $l+d$. After creating the string between external charges, we approximate the real-time dynamics as a sequence of small time steps $\Delta t$, where we use a first-order Trotter decomposition of the time-evolution operator corresponding to
\begin{equation}
    e^{-i\Delta tH}\approx e^{-i\Delta tH_\text{el}} e^{-i\Delta tH_m}  e^{-i\Delta tH_\text{kin}} 
\end{equation}
for each step.

\begin{figure}[htp!]
    \centering
    \includegraphics[width=\linewidth]{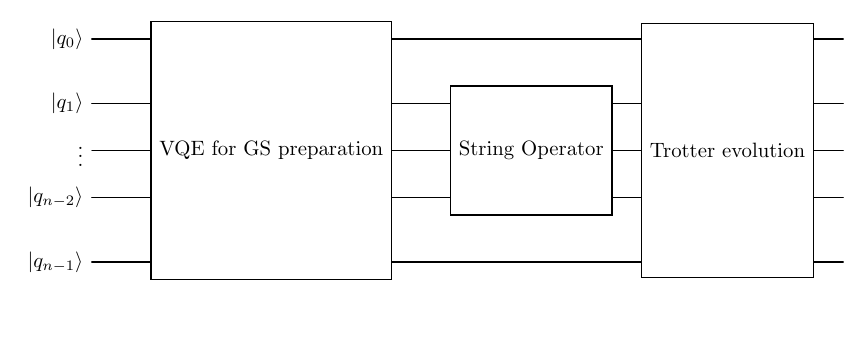}    
    \caption{Diagrammatic presentation of the overall circuit structure used for simulating real-time evolution of string breaking. First we prepare the ground state of the system using the final parameters obtained  from the VQE. Subsequently, we apply a string of Pauli $Z$-matrices to create a string between a pair of external charges. Finally, we evolve this state in time using a first-order Trotter decomposition of the time evolution operator.}
    \label{fig:schema-time_evol}
\end{figure}

Note that the circuit implementing a single time step essentially corresponds to one layer of our VQE Ansatz circuit (see Fig.~\ref{fig:circuit_no1stlayer}), with the only difference that the parameters are now no longer variational but are determined by the Hamiltonian parameters and the time step. Figure~\ref{fig:schema-time_evol} shows a diagrammatic presentation of the entire procedure used to simulated dynamical aspects of string breaking.

\section{Numerical results \label{sec:numerical_results}}
In this section, we present the findings of our investigation of string breaking on the lattice, which we conduct in two stages. Firstly, we focus on the static potential, and demonstrate that it allows us to reliably determine the parameter regions in which we expect string-breaking to occur. Studying the site-resolved modular charge and electric field configuration in the ground state, we show that these observables provide a clear signal for string breaking. Moreover, we demonstrate that our VQE Ansatz combined with proper initialization allows for preparing the ground state of the model for both the absence and presence of static charges with high fidelity. Secondly, we explore the dynamical aspects of string breaking by superimposing a string between  external charges in the ground state  and evolving the resulting configuration in time. 

\subsection{Static potential}
To study the static potential,  we use MPS to determine the static potential of the theory, also for system sizes that cannot be easily reached with current quantum hardware, and show that we can reliably determine the breaking point of the string. Subsequently, we show that our VQE Ansatz is capable of reproducing the static potential on current quantum hardware.

\subsubsection{Tensor Network simulations}
\begin{figure}[htp!]
    \centering    
    \includegraphics[width=0.95\linewidth]{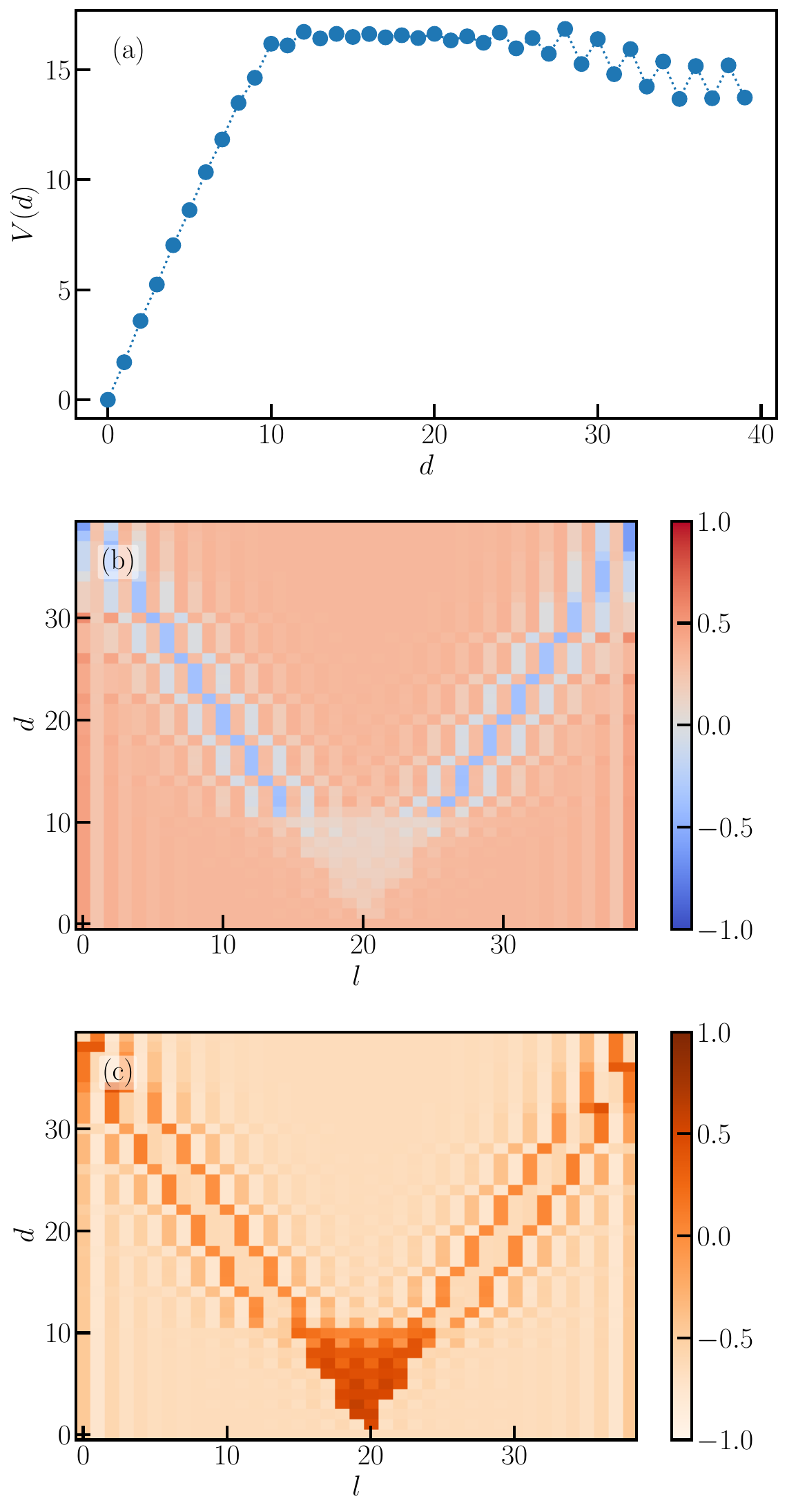}
    \caption{(a) Static potential for a system with $L=40$, $J=30.0$, $m=3.0$, and $\varepsilon=1.5$. (b) Site resolved expected value of $p_l$ indicated by the color for varying distance and (c) site resolved expected value $X_{l,l+1}$. A distance of $d=0$ corresponds to the ground state in the sector without static charges, whereas values of $d\geq 1$ represent configurations in the presence of a pair of static charges.}
    \label{fig:static_string_breaking_tn}
\end{figure}

To explore string-breaking in a static manner, we use MPS to obtain the ground state of the theory. First, we compute the ground state of the theory in the sector with vanishing external charges, before we repeat the calculation in the presence of a pair of static charges at a distance $d$. To avoid finite size effects, we place the pair of static charges approximately in the middle of the system. From these simulations we can extract the static potential, as well as the site-resolved configuration of the modular charges and the electric field. Figure~\ref{fig:static_string_breaking_tn} shows the results for a system with  $L=40$, corresponding to 79 qubits, $J=30.0$, $m=3.0$, and $\varepsilon=1.5$.
From the results on the static potential shown in Fig.~\ref{fig:static_string_breaking_tn}(a) one can clearly see that $V(d)$ shows two distinct regions. For small distances, we observe a linear increase, as expected for the regime in which a flux string is present in the system. In contrast, for distances $d\gtrsim 11$ the static potential flattens, thus indicating that the flux string eventually breaks. Note that $V(d)$ does not become completely flat after reaching the breaking point of the string. This can be attributed to two reasons i) Since we are working with a staggered formulation even distances correspond to the static charges being located between two (anti-)matter sites having the same positive or negative sign in the staggered mass term, whereas odd distances have  one static charge on a matter site and one on an antimatter site, leading to the observed oscillatory behavior  in the values within the plateau region. ii) As we get close to the boundaries, finite-size effect become relevant, which tend to enhance the difference between odd and even distances as well as to decrease the values in the  the plateau region of  the static potential.

The picture obtained from the static potential is corroborated by the electric field shown in Fig.~\ref{fig:static_string_breaking_tn}(c). For the ground state in the absence of static charges the expected value of $X_{l,l+1}$ is negative. As soon as two static charges are present,  we observe a region at the center of the system where the gauge links are excited manifesting itself in a positive value of $X_{l,l+1}$. The flux tube stretches with increasing separation of the charges, up to a distance $d\geq 11$, at which it eventually vanishes, signaling the onset of string-breaking. At the breaking point, we also observe the creation of charges at the end of the flux tube (see Fig.~\ref{fig:static_string_breaking_tn}(b)), reflected by the negative value of $p_l$. This indicates the formation of two mesons between an external charge and a matter site at the edge of the string. 

\begin{figure}[htp!]
    \centering
    \includegraphics[width=0.95\linewidth]{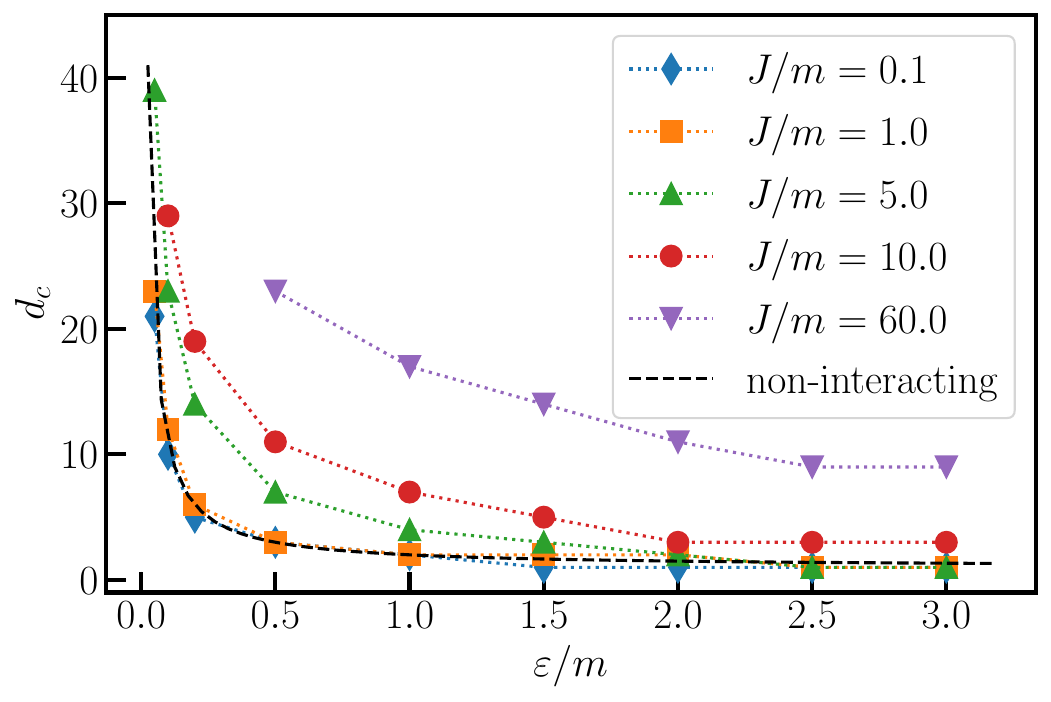}
    \caption{Critical distance $d_c$ at which the string breaks for a system with $L=40$ as a function of $\varepsilon/m$   extracted from the static potential and site-resolved electric field configurations from our  simulations for various values of $J/m$. Symbols connected with dotted lines  correspond to the same value of $J/m$  as a guide for the eye. The black dashed line shows the theoretical prediction for the noninteracting case from Eq.~\eqref{eq:critical_distance_sc}.}
    \label{fig:critical_distances}
\end{figure}

Since  we can always multiply the Hamiltonian in Eq.~\eqref{eq:Z2_Hamiltonian_fermionic} with a nonzero constant without changing the physics, we measure $J$ and $\varepsilon$ in units of mass, without loss of generality. In Fig.~\ref{fig:critical_distances}, we show the breaking point for a system with $L=40$ as a function of $\varepsilon/m$ for various values of hopping parameter $J/m$. 
As expected, for small values of $J/m \ll 1$, the system essentially follows the noninteracting prediction from Eq.~\eqref{eq:critical_distance_sc} for the entire range of $\varepsilon/m$. This behavior persists up to $J/m=1.0$, for which we still do not observe noticeable deviations from the noninteracting limit. Our data show some deviation from  the noninteracting regime when increasing the hopping parameter to $J/m=5.0$. However, for values of $\varepsilon/m\geq 2.0$ we again approach the noninteracting limit. Increasing the hopping parameter further, these deviations become more pronounced. In particular, for $J/m=60.0$ we  no longer observe string breaking for $\varepsilon/m < 0.5$. In general, comparing data for a fixed $\varepsilon/m$, we observe  an increase of the critical distance with increasing values of $J/m$.  While such a behavior  might seem counterintuitive at first, this can be explained by the fact that the interaction lowers the relative excess energy of the string compared to the ground state energy without static charges. Thus, for increasing values of $J$ the flux string survives up to larger distances before it eventually breaks.

\subsubsection{VQE approach\label{subsec:string_breaking_vqe}}

In order to demonstrate that the VQE we are using is suitable for capturing the static properties of string breaking, and to obtain the ground state for subsequent dynamical simulations, we first study the static potential following a procedure analogous to our MPS approach. We benchmark the performance of our VQE protocol in two ways. First, we simulate the VQE classically assuming a perfect quantum computer performing an infinite number of measurements, i.e.\ simulating the VQE using the state vector. Second, we introduce shot noise by using 10,000 measurements per iteration. In both cases, the VQE optimization uses  to 10,000 iterations. 
The parameters in the circuit are initialized with small values drawn from a uniform distribution in the interval $[0,10^{-4}]$, ensuring that the initial state is close to the solution for the noninteracting limit. We also examine different Ansatz depths $n_l$ and compare the final results to the exact solution via exact diagonalization. Finally, we prepare the state obtained at the end of the ideal VQE on a quantum device and measure the energy, and the site-resolved charge and link configurations. Using zero-noise extrapolation (ZNE) and Twirled Readout Error eXtinction (TREX)~\cite{PhysRevA.105.032620} to mitigate errors, we show that our Ansatz circuit is suitable for current quantum hardware (for details on error mitigation see Appendix~\ref{section:ZNE}).

In Fig.~\ref{fig:static_string_breaking_1}, the static potential is shown for a system with $L=10$, which corresponds to 19 qubits, computed with different methods. 
\begin{figure}[htp!]
    \centering
    \includegraphics[width=\linewidth]{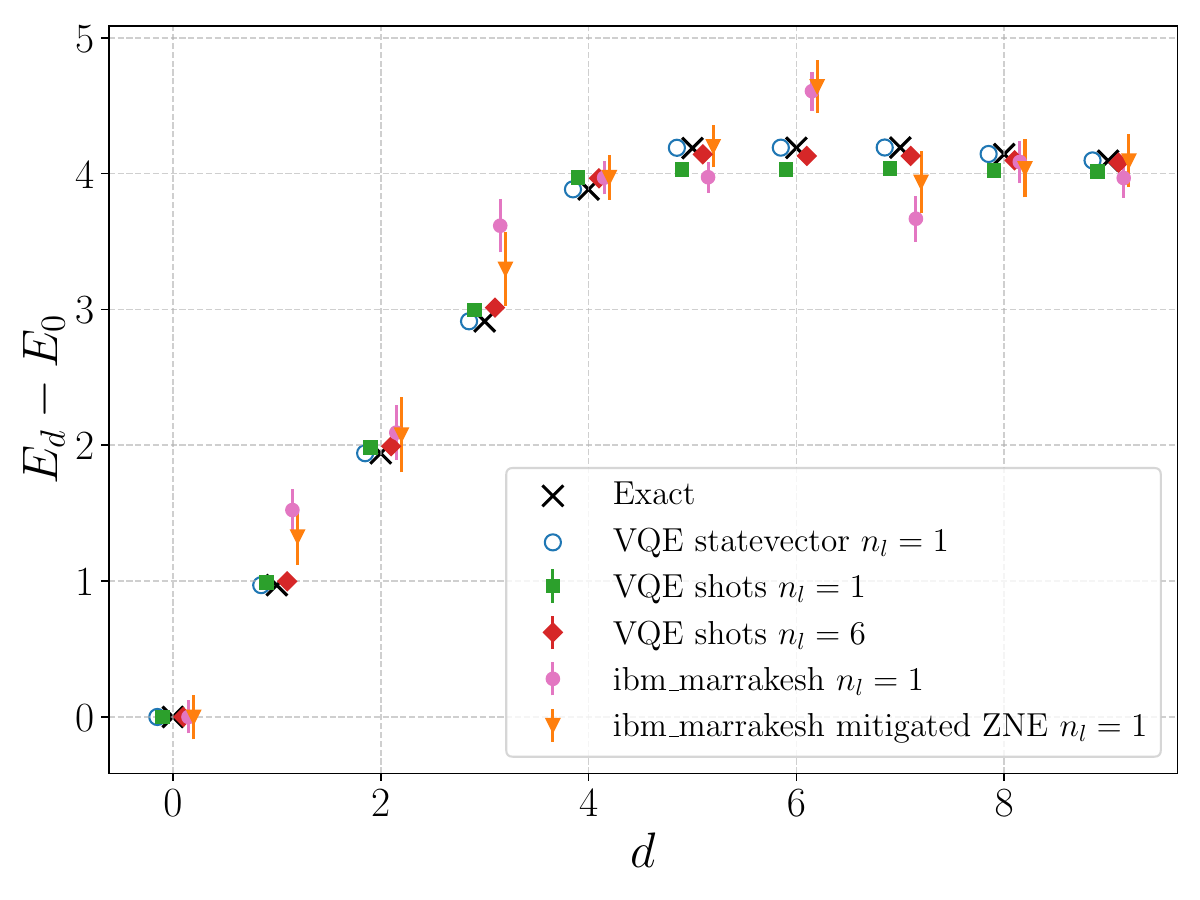}
    \caption{Static potential for $J = 0.5$, $m = 2.0$, $\varepsilon = 0.5$, and system size $L = 10$ (corresponding to 19 qubits). 
    Black crosses indicate results obtained from exact diagonalization. 
    Blue circles represent VQE results obtained via classical simulation using the state vector and a single-layer ansätze $(n_l =1)$. Colored markers with error bars show VQE results simulated with 10,000 measurements per iteration, averaged over 100 independent runs corresponding to an $n_l=1$ (green) and $n_l=6$ (red) Ansätze respectively. The error bars in this case are computed as $\Delta E = \sqrt{ \sigma(E_d)^2+\sigma (E_0)^2}$, where $\sigma$  denotes the standard deviation over these 100 runs. Pink circles represent unmitigated results from a single run on the quantum device \texttt{ibm\_marrakesh}, using the final parameters obtained from classical state vector simulations. The device error bars are computed as $\Delta E = \sqrt{ (\Delta E_d)^2 + (\Delta E_0)^2 }$ where $\Delta E_d$ is the hardware noise at distance $d$.
    Orange triangles show the same data after applying linear zero-noise extrapolation (ZNE) for error mitigation. The error is computed in a similar manner as before using the ZNE-extrapolated error of the energy $\Delta E^\text{ZNE}_d$. Different markers are slightly offset horizontally for better visibility.    
    }
    \label{fig:static_string_breaking_1}
\end{figure}

Examining first on the classically simulated VQE using the state vector, we observe excellent agreement with the exact diagonalization results, both in the linearly rising regime and in the plateau region. Notably, our Ansatz is able to capture the relevant physics with just a single layer in the parameter regime studied, achieving fidelities of at least $99.88\%$ (see Appendix~\ref{app:circuit_initialization} for details). The minor deviations are due to VQE being a variational methdod, finite numerical precision, and the convergence behavior of the classical optimization routine, and are on the order of $10^{-3}$ to $10^{-5}$.

Introducing shot noise by using a finite number of measurements in each iteration of the classical simulation of the VQE only moderately affects the convergence, which highlights the effectiveness of our chosen Ansatz. Increasing the number of layers leads to minimal improvements in convergence. Each data point represents the average over 100 independent runs, where the uncertainty in the ground-state energy is quantified by the standard deviation across these runs. The error bars shown in Fig.~\ref{fig:static_string_breaking_1} are then computed via standard error propagation, taking into account the uncertainties of both the ground-state energy with and without static charges.

\begin{figure}[hbp!]
    \centering
    \includegraphics[width=\linewidth]{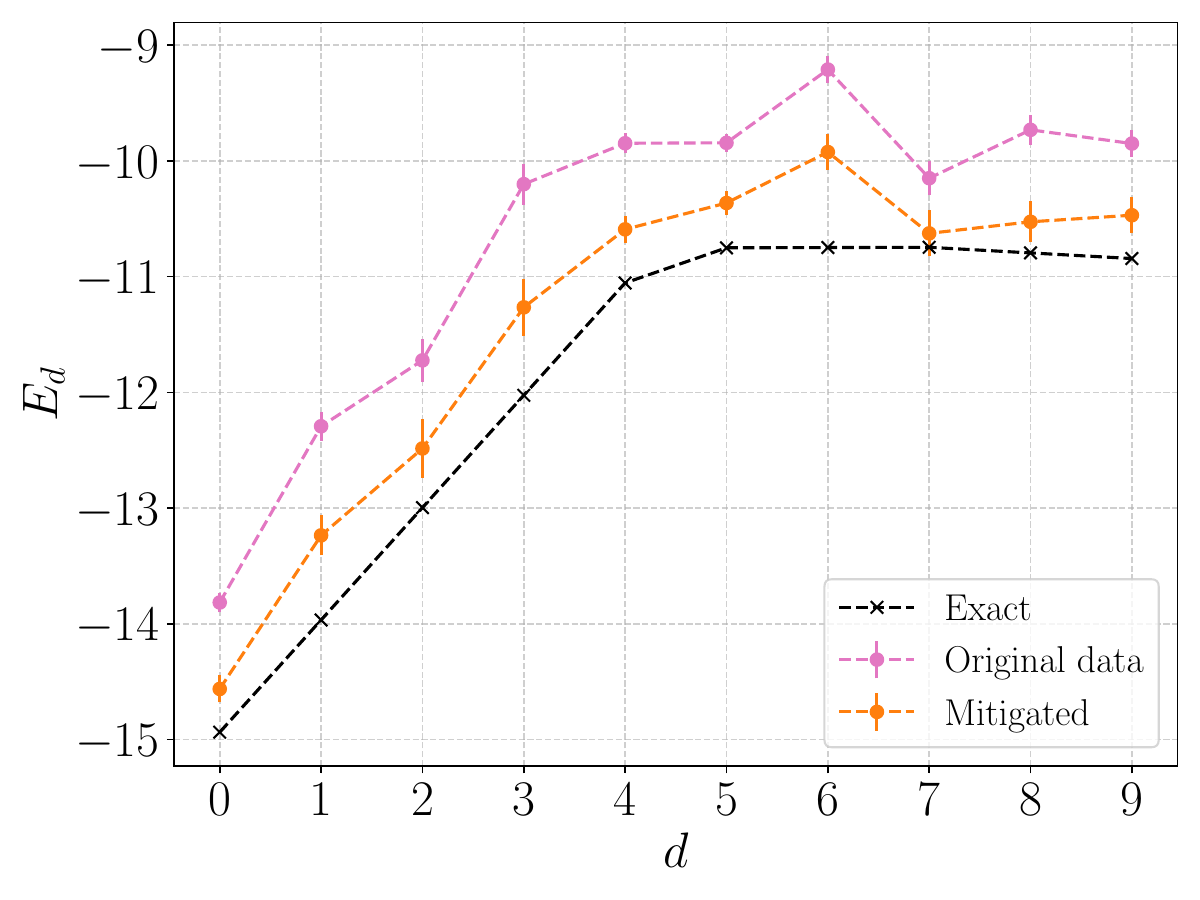}
    \caption{
    Ground state energy as a function of the separation distance \( d \) between static charges for \( J = 0.5 \), \( m = 2.0 \), \( \varepsilon = 0.5 \), and system size \( L = 10 \), using a single-layer Ansatz (19 qubits). 
    We present the hardware results from \texttt{ibm\_marrakesh} before ZNE (pink), the same data after applying ZNE (orange), and the expected values from exact diagnonalization(black). 
    Unlike in the static potential plots where \( E_d - E_0 \) is shown, this figure displays the energy values directly, allowing for a clearer interpretation of the effect of error mitigation.    
    }
    \label{fig:energy_vs_distance}
\end{figure}

To demonstrate that the VQE can also be carried out on current quantum hardware, we prepare the state for the final parameters obtained for the classical simulation using the state vector on quantum hardware\footnote{In principle, the entire VQE can be run on quantum hardware. However, this requires quite a significant amount of computing time on hardware, thus we restrict ourselves to demonstrating state preparation as required for every iteration.}. Our data for $V(d)$ obtained on \texttt{ibm\_marrakesh} are shown in Fig.~\ref{fig:static_string_breaking_1}. Our data reproduces the behavior of the exact result for the static potential well, and we can reliably distinguish the two different regions in $V(d)$, even without applying ZNE. Applying ZNE generally improves the results, except for $d=6$ where the ZNE is rendering the results slightly worse. To elucidate the effect of ZNE on the energy in depth, we show in Fig.~\ref{fig:energy_vs_distance} a comparison of ground state energy as a function of distance of the static charges. Looking at the ground state energy instead of the static potential, we observe that ZNE consistently improves the results, and the final energies are closer to the exact solution. While the unmitigated data for the energy is noticeably above the exact data, the static potential corresponds to the difference $E_0 - E_d$. Hence, this offset is reduced when looking at $V(d)$, thus explaining the small deviations in the static potential before error mitigation. Moreover, we see that although ZNE reduces the error for each energy value, the reduction is not exactly the same for $E_0$ and $E_d$, leading to the result that the mitigated data on the energy difference shown in Fig.~\ref{fig:static_string_breaking_1} to sometimes look slightly worse than the unmitigated one.

\begin{figure}[htp!]
    \includegraphics[width=\linewidth]{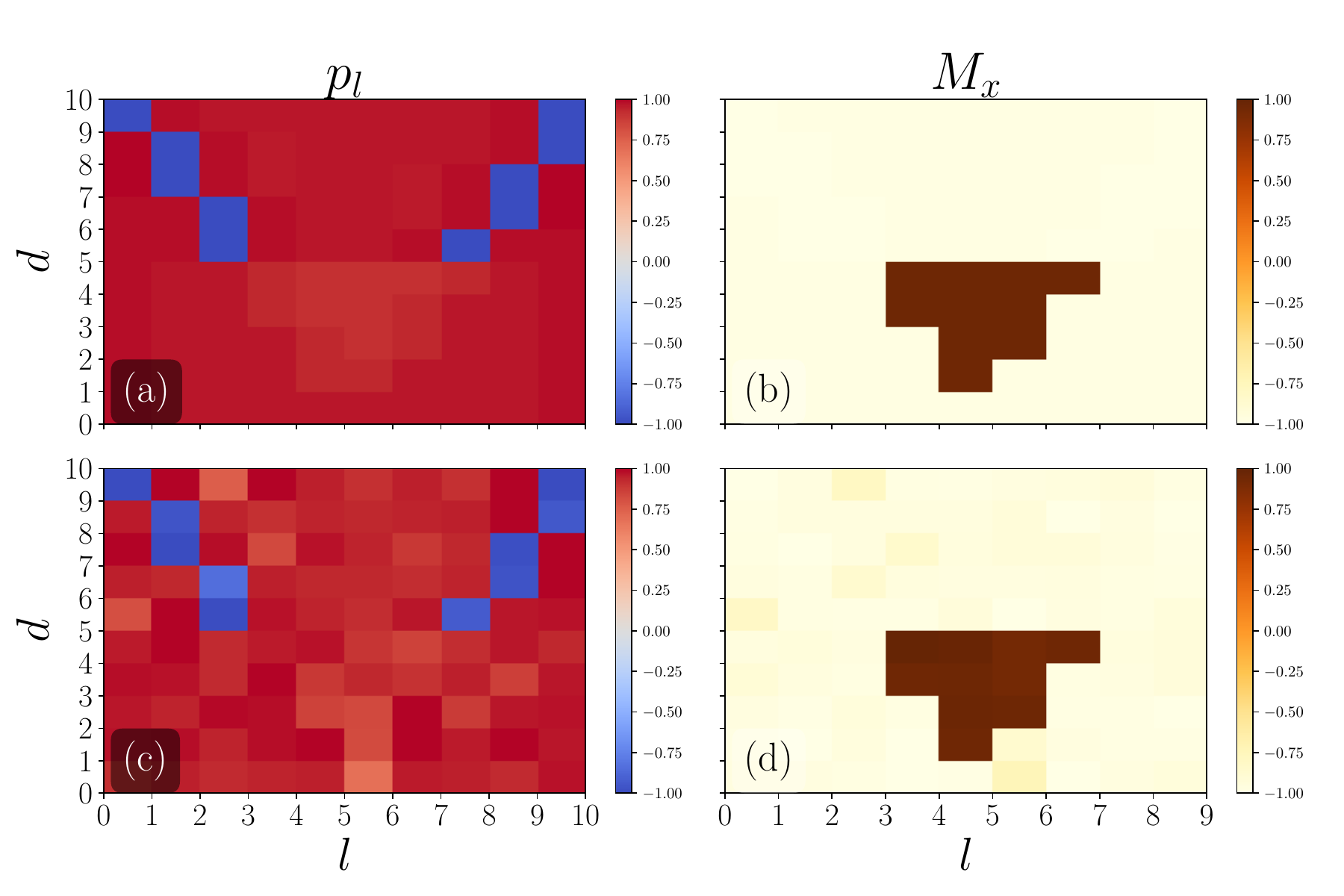}
    \caption{Site-resolved configuration for various distances of the static charges with $J=0.5$, $m=2.0$, $\varepsilon = 0.5$ and $L=10$, corresponding to 19 qubits. The left panels show the expected value of the modular charges, $p_l$, the right panels display the expected value of the Pauli-$X$ matrix on the gauge links. The top row shows exact results obtained from state vector simulation, while the bottom row depicts results measured on \texttt{ibm\_marrakesh} after error mitigation. 
    }
    \label{fig:colormap_static_breaking_N19}
\end{figure}

Similar to the MPS simulations, in Fig.~\ref{fig:colormap_static_breaking_N19} we also present the site-resolved configuration of the modular charges $p_l$ and gauge links after ZNE. 
Comparing the upper row, which presents the ideal results obtained using state vector simulations, to the bottom one, which depicts the results from \texttt{ibm\_marrakesh} after ZNE, we generally observe good agreement. In particular, we can clearly see that the linear region in the static potential (c.f.\ Fig.~\ref{fig:static_string_breaking_1}) corresponds to having a flux tube in Fig.~\ref{fig:colormap_static_breaking_N19}(d), which eventually vanishes as the distance is increased beyond the point where a plateau in $V(d)$ starts forming. From this point on the charge configuration in Fig.~\ref{fig:colormap_static_breaking_N19}(c) shows two pronounced sites with a modular charge close to -1, thus indicating again the breaking of the flux tube.

\begin{figure}[htp!]
    \includegraphics[width=\linewidth]{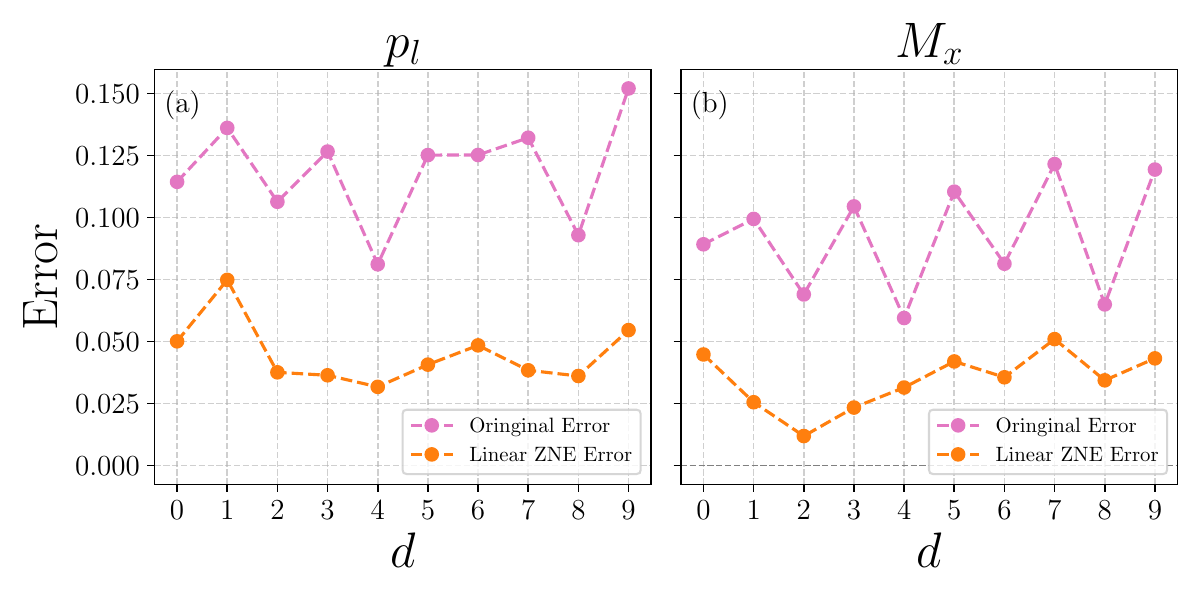}
    \caption{Average absolute error (over lattice sites/links) of the results obtained on \texttt{ibm\_marrakesh} for various distances with $J=0.5$, $m=2.0$, $\varepsilon = 0.5$ and $L=10$, corresponding to 19 qubits. The two different lines represent the errors of the original results without any error mitigation (pink circles) and  after ZNE using a linear fit function (orange circles).
    }
    \label{fig:abs_error_colormap_static_breaking_N19}
\end{figure}

Additionally, we examine the effect of the ZNE on these observables in greater detail by subtracting the experimental data in Figs.~\ref{fig:colormap_static_breaking_N19}(c) and (d) from the exact results of Figs.~\ref{fig:colormap_static_breaking_N19}(a) and (b). The average of this absolute error for both the modular charge and the gauge link configuration is shown in Fig.~\ref{fig:abs_error_colormap_static_breaking_N19}.
For both observables, we see that applying ZNE decreases the error by more than  a factor of 2 (see Appendix~\ref{section:ZNE} for details). Note that the average absolute error is quite small compared to the values of the modular charge and electric field in Fig.~\ref{fig:colormap_static_breaking_N19}.

Overall, our data show that a VQE with moderately deep Ansatz circuit is able to capture the relevant physics of the model for both the absence and the presence of static charges. Moreover, the configurations of the gauge links and the modular charges together with the static potential allow for identifying the regimes in which string breaking occurs. In addition, our data on quantum hardware demonstrate that the corresponding states can be reliably prepared on current quantum device.

\subsection{Real-time dynamics of string breaking\label{subsec:real_time_dynamics_results}}
Starting from the ground state, we now turn to dynamical simulations of string breaking. To create a flux string excitation starting from the ground state, we proceed as outlined in Sec.~\ref{subsec:qc_evo} and apply the string operator from Eq.~\eqref{eq:string_operator} without the operators acting on the matter sites at the end. Subsequently, we evolve the resulting state in time and monitor the site resolved configuration of the matter sites and gauge links.

\subsubsection{Tensor Network simulations}
In order to obtain an idea how the string behaves over time in both, regimes where we expect the flux tube to break and where no string breaking occurs, we first study the problem with MPS. To simulate the real-time evolution, we use the time-evolution block decimation approach with an odd-even decomposition of the time evolution operator~\cite{Vidal2003,Schollwoeck2011}. While this introduces some Trotter error, we checked that our time step of $\Delta t=0.0025$ is small enough to avoid noticeable effects throughout the entire time we simulate. In particular, for all the cases we study, the relative change in energy over the entire time we simulate is smaller than $10^{-3}$.

\begin{figure}[htp!]
    \includegraphics[width=\linewidth]{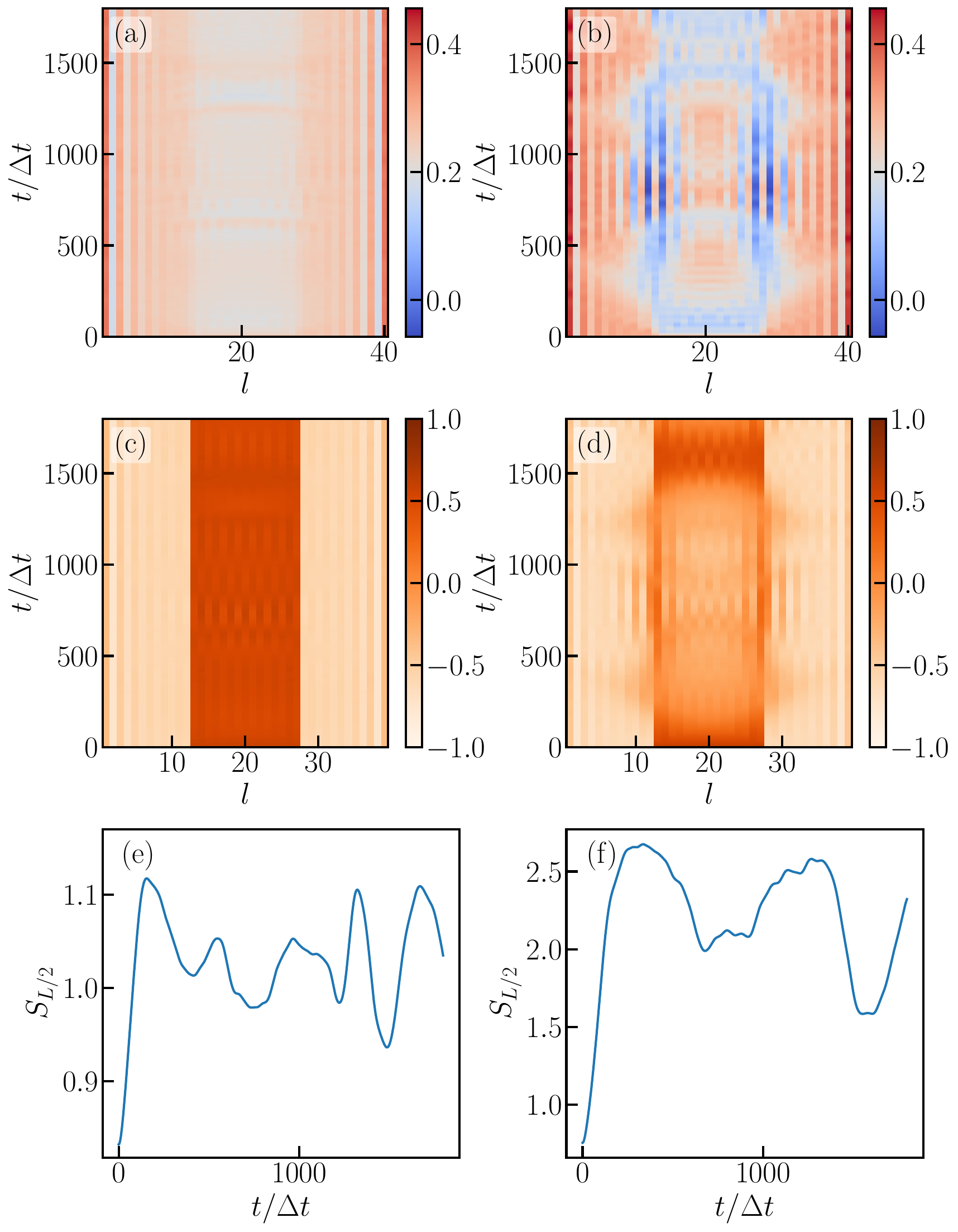}
    \caption{Site-resolved modular charges (first row), configuration of the gauge links (second row) and von Neumann entropy obtained for bipartitioning the system in the center (third row) as a function of time for the evolution of a flux tube of length $d=15$  for $L=40$, $J=30.0$, $m=3.0$ and $\varepsilon=0.25$ (left column) and $\varepsilon=1.0$ (right column).}
    \label{fig:mps_evo_N40}
\end{figure}

Figure~\ref{fig:mps_evo_N40} shows our results for a system with $J=30.0$, $m=3.0$ and $L=40$, corresponding to 79 qubits. 
Examinining the left column, which represents the case of $\varepsilon=0.25$, we see that the flux string is essentially stable over the entire time we simulate. The site-resolved modular charge configuration (c.f.\ Fig.~\ref{fig:mps_evo_N40}(a)) does not show significant changes over time, and also the site-resolved configuration of the gauge links is essentially constant and maintains the string configuration over the entire time of the simulation, as revealed in Fig.~\ref{fig:mps_evo_N40}(c). The minor fluctuations observed in these two observables can be attributed to the fact that the initial state is not an eigenstate of the Hamiltonian. This is also evident in the von Neumann entropy, as shown in Fig.~\ref{fig:mps_evo_N40}(e), which initially exhibits a minor increase before oscillating around a moderate value of 1.

In contrast, for a larger value of $\varepsilon=1.0$, the electric energy contribution is large enough to render string breaking favorable, which can be clearly seen in the right column of Fig.~\ref{fig:mps_evo_N40}. In particular, we observe that at the beginning of the dynamics charges are formed in inside the flux string (c.f.\ blue region at the bottom of Fig.~\ref{fig:mps_evo_N40}(b)). These eventually move towards the location of the static charges and lead to the breaking of the string beginning around $t/\Delta t \approx 250$, which can be clearly seen in Fig.~\ref{fig:mps_evo_N40}(d) as an emerging depletion in the electric flux in the region the string started. Interestingly, the von Neumann entropy, $S_{L/2}$, shown in Fig.~\ref{fig:mps_evo_N40}(f), exhibits a clear peak at the point where the string breaks, and the maximum value is much larger than in the previous case. Subsequently, $S_{L/2}$ decreases again until  $t/\Delta t \approx 670$. At this time, we observe a clustering of charges and high electric flux only at the edge of the initial string region, indicating the formation of mesons between the external charges and the matter sites at the edge of the string.

Following the initial breaking of the flux string, the von Neumann entropy starts to grow again around $t/\Delta t \approx 900$ before eventually reaching a peak, which is then followed by a revival of the string around $t/\Delta t \approx 1400$ (c.f.\ Fig.~\ref{fig:mps_evo_N40}(d)). After the revival, we again observe a picture similar to the initial phase of the evolution, where charges are created, the string begins to break and we observe a clear drop in the entropy, before the entropy is again increasing. These oscillation are similar to the behavior observed for a flux string between static charges in the Ising spin chain and a U(1) gauge theory~\cite{Sala2018, De:2024smi}.

\subsubsection{Trotterized time evolution on quantum hardware}

The classical simulation of the time evolution in the previous subsection demonstrated that while the initial value of the entropy was similar for parameter sets we studied, it showed a considerable growth over time in the case the flux string breaks. This motives using quantum hardware to simulate the out-of-equilibrium dynamics of string-breaking. To investigate the real-time dynamics of a flux string on a quantum device, we follow the methodology outlined in Sec.~\ref{subsec:qc_evo}, using a trotterization of the time evolution operator. This requires choosing a sufficiently small time step, $\Delta t$, to minimize errors from the Trotter approximation while ensuring accurate simulation results. At the same time, $\Delta t$ must be large enough to allow meaningful observation of string breaking dynamics within a number of time steps that remains feasible given circuit depth constraints, as noise in current quantum hardware limits the depth of circuits that can be executed meaningfully.
For  $J=0.5$, $m=0.1$, and  $\varepsilon=0.15$,  used in the Hamiltonian, a time step  of $\Delta t=0.5$ balances these competing factors, for up to 10 time steps sufficiently well. Note that for our choice of parameters, Fig.~\ref{fig:critical_distances} suggests that we are away from the noninteracting limit. Moreover, from a calculation of the static potential and the gauge link configuration, we see that a flux tube of length length $d=7$ is not expected to be stable in this regime. Thus, preparing the ground state via VQE and creating a flux tube  allows us to observe the dynamics of string breaking.

\begin{figure}[htp!]
    \centering
    \includegraphics[width=\linewidth]{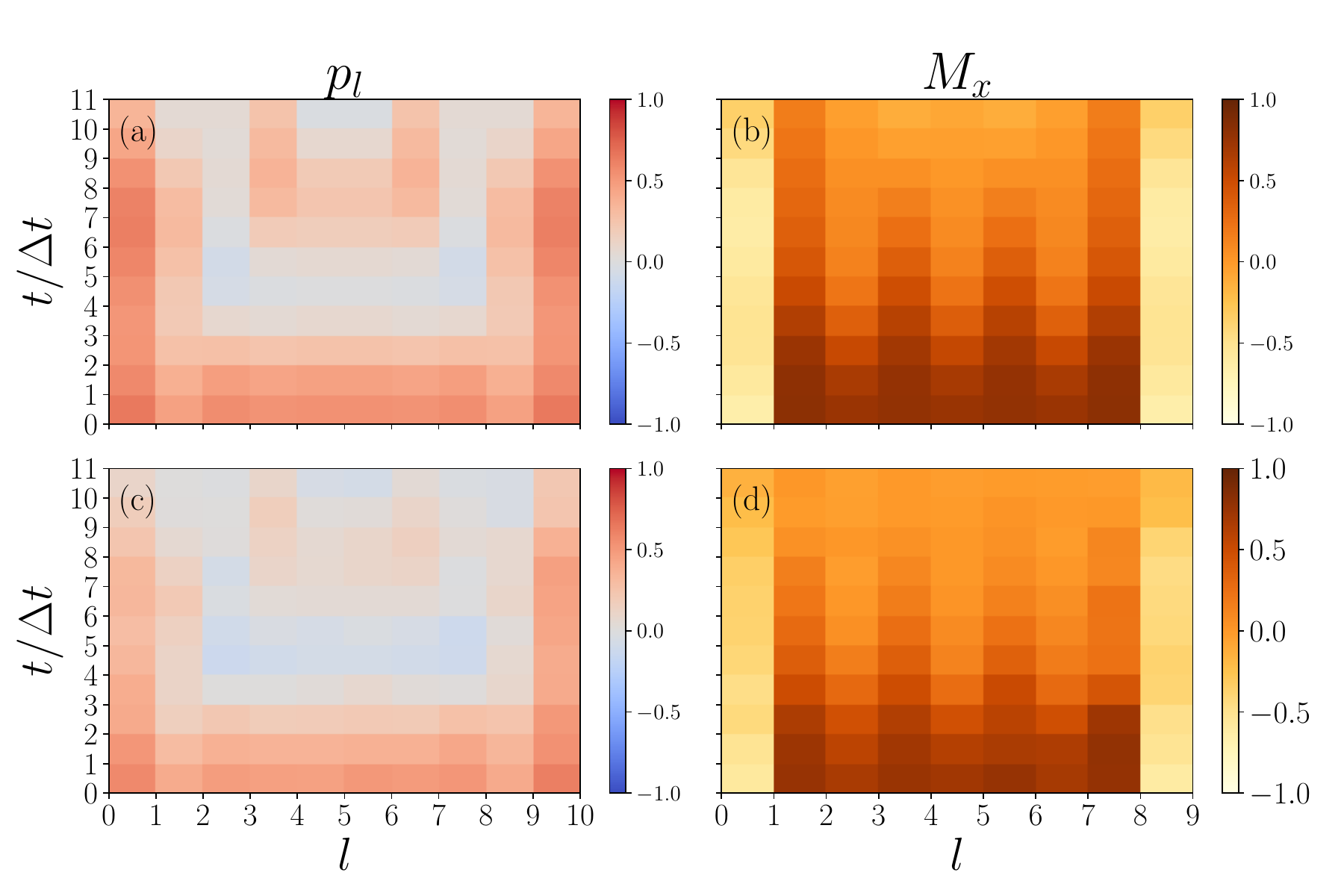}
    \caption{The top row shows exact trotter time evolution results obtained from state vector simulation, while the bottom row depicts results from the \texttt{ibmq\_fez} device, corrected with zero-noise extrapolation (ZNE) using gate folding,  $L=10$ matter sites (19 qubits) chain with  $J=0.5$, $m=0.1$ and $\varepsilon=0.15$.   A linear fit is used for the extrapolation to obtain the mitigated expectation values. Device simulations are performed with time steps $\Delta t=0.5$ until  time $T = 5$.}
    \label{fig:trotter_N19_J05_m01_e015}
\end{figure}

Figure~\ref{fig:trotter_N19_J05_m01_e015} shows our data for the  site-resolved modular charge $p_l$ (left panels) and the electric field on the gauge links, $M_x$, (right panels) as functions of time. The top row displays the ideal results obtained from a state vector simulations, serving as a noiseless benchmark. The bottom row shows the results obtained from the \texttt{ibm\_fez} quantum device after applying the ZNE.

Comparing the exact results with the ones from the quantum hardware, we observe very good agreement, with minor deviations only occurring after several time steps. Similarly to our previous MPS simulations, we observe the characteristic behavior that charges are initially generated within the flux tube, which then gradually redistribute towards the static charges at the edges of the flux tube. This ultimately results in the breaking of the initial string after approximately four time steps.

\begin{figure}[htp!]
    \centering
    \includegraphics[width=\linewidth]{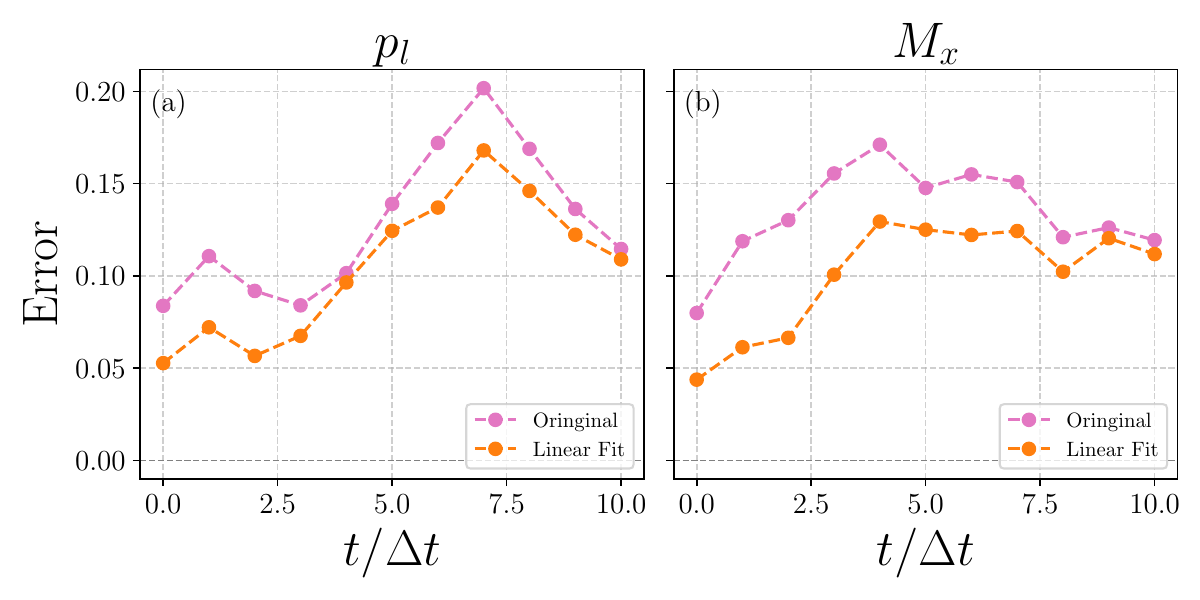}
    \caption{Average absolute error (over lattice sites/links) of the results in Fig.~\ref{fig:trotter_N19_J05_m01_e015} obtained on \texttt{ibm\_fez} for $J=0.5$, $m=0.1$ and $\varepsilon=0.15$ as a function of time. The notation is the same as in Fig. \ref{fig:abs_error_colormap_static_breaking_N19}.}
    \label{fig:error_trotter_N19_J05_m01_e015}
\end{figure}

Again, we investigate the effect of the ZNE by looking at the absolute average error between the ideal results and the experimental results from the quantum device. The data for the modular charges and the configuration of the gauge links are presented in Fig.~\ref{fig:error_trotter_N19_J05_m01_e015}.
In general, we see that for both observables, ZNE noticeably improves the error. While at the beginning both linear and exponential fits for the ZNE show comparable results, we observe that for later times the exponential fit seems to provide better results. 

\section{Conclusions \label{sec:conclusions}}

In this work, we explore string breaking in a (1+1)-dimensional $\Zt$ lattice gauge theory coupled to Kogut-Susskind staggered fermions, studying both static and dynamical aspects of the phenomenon. Using TNs, we  investigate the static aspects of string breaking over a wide range of parameters, demonstrating that the static potential between charges, as well as the corresponding site-resolved matter and gauge link configurations, allow for reliably identifying the onset of string breaking. In addition, we  develop a VQE protocol that is able to study the static aspects of string breaking and have benchmarked our Ansatz using both classical simulations and carrying out inference runs on real quantum hardware with up to 19 qubits. Our results demonstrate that the VQE proposed is able to capture the physics of the system with low-depth circuits that are suitable for current quantum hardware.

Furthermore, we  study the real-time dynamics of a flux between static external charges. Carrying out TN simulations in a regime where string breaking occurs, we demonstrate that site-resolved configuration of the matter sites and gauge links over time allows to gain insight into the breaking process of the flux tube. In particular, we  observe that in an initial step charges are generated inside the string, which eventually redistribute towards the external charges before eventually breaking the flux tube and lowering the electric field in the center. Moreover, after the initial breaking the flux tube, string formation repeatedly reappears. The breaking process is accompanied with a significant production of entanglement entropy, which motivates the use of quantum hardware. 

Using a trotterization of the time evolution operator, we  also demonstrate the breaking of a flux tube between static charges during the evolution on real quantum hardware for a system of 19 qubits. Using ZNE to mitigate errors, our data from the quantum device are in very good agreement with the exact results. In particular, we are able to observe the production of charges inside the flux tube at the beginning of the evolution and their redistribution towards the static charges, eventually leading to  string breaking.

The results obtained in this work demonstrate  the potential of quantum algorithms for simulating dynamical properties of gauge theories, thus allowing to access regimes where conventional methods are not applicable. Future work will extend these techniques to larger systems, explore alternative Ans\"atze for improved efficiency, and implement advanced error mitigation strategies to enhance performance on real quantum hardware. This study serves as one of the necessary first steps towards studying further out-of-equilibrium problems of gauge theories, providing insights into non-perturbative phenomena beyond the capabilities of conventional lattice methods.

\acknowledgments
We thank Dr.\ Yahui Chai for helpful discussions regarding error mitigation on IBM's quantum hardware.
G.\ P.\ received financial support from the Cyprus  Research and Innovation Foundation under contract number POST-DOC/0718/0100, from projects 3D-nucleon and NextQCD, co-funded by the European Regional Development Fund and the Republic of Cyprus through the Research and Innovation Foundation with contract id EXCELLENCE/0421/0043 and EXCELLENCE/0918/0129 respectively and the University internal project PDFs-LQCD.
C.\ A.\ is partially supported by the  EJD project AQTIVATE funded under  the MSCA of the European Union, grant agreement no.\ 101072344.
A.\ A.\ is supported by the European Union's Horizon 2020 research and innovation programme under grant agreement No 810660.
This work is supported with funds from the Ministry of Science, Research and Culture of the State of Brandenburg within the Center for Quantum Technology and Applications (CQTA). 
\begin{center}
    \includegraphics[width = 0.05\textwidth]{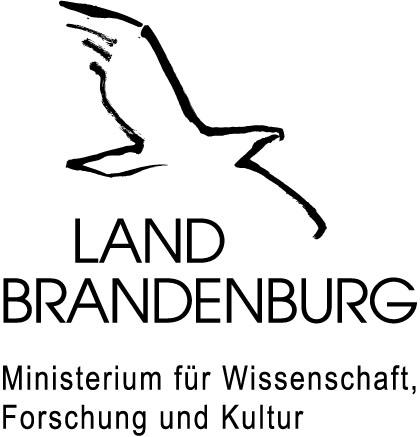}
\end{center}

\appendix

\section{Circuit initialization and fidelity of the VQE\label{app:circuit_initialization}}

\begin{figure*}[htp!]
    \centering
    \includegraphics[width=\linewidth]{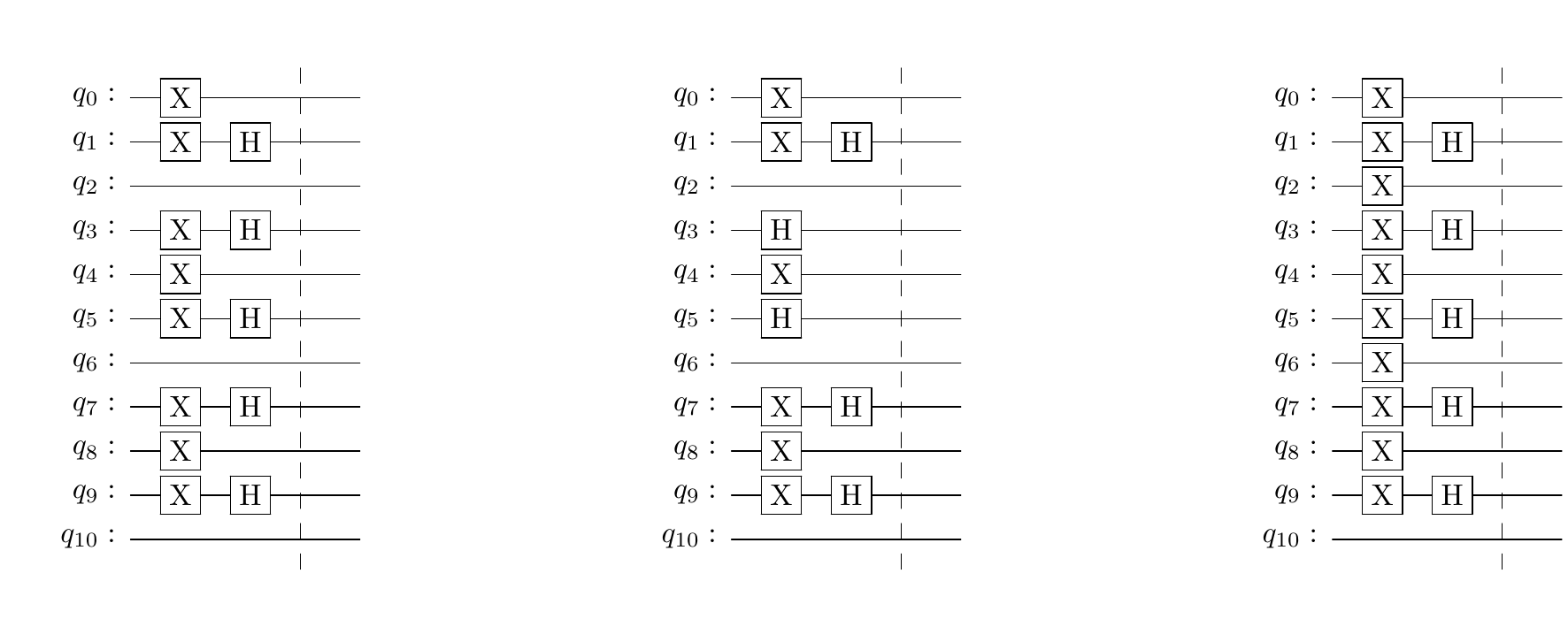}
    \caption{Circuits for preparing the initial states. Left panel:  ground state with no external charges corresponding to $\ket{1}\ket{-}\ket{0}\ket{-}\ket{1}\ket{-}\ket{0}\ket{-}\ket{1}\ket{-}\ket{0}$. Middle panel: flux tube between external charges located at matter sites 3 and 5 $\ket{1}\ket{-}\ket{0}\ket{+}\ket{1}\ket{+}\ket{0}\ket{-}\ket{1}\ket{-}\ket{0}$. Right panel: configuration with external charges at sites 3 and 5 with with dynamical charges at these sites shielding them and no flux tube, corresponding to $\ket{1}\ket{-}\ket{1}\ket{-}\ket{1}\ket{-}\ket{1}\ket{-}\ket{1}\ket{-}\ket{0}$.    
    }
    \label{fig:new_initialization_circuit}
\end{figure*}

As we discussed in the main text, we choose $\ket{\psi_\text{init}}$ to be either the ground state without static charges, a state with a flux tube or a state with two mesons in the noninteracting limit, $J\to 0$. These states can be easily prepared from the state $\ket{00\dots 0}$, in which the IBM quantum devices are initialized, with just single-qubit gates. Explicit examples of the corresponding circuits for $L=6$, corresponding to 11 qubits are shown in in Fig.~\ref{fig:new_initialization_circuit}.

\begin{figure}[htp!]
    \centering
    \includegraphics[width=\linewidth]{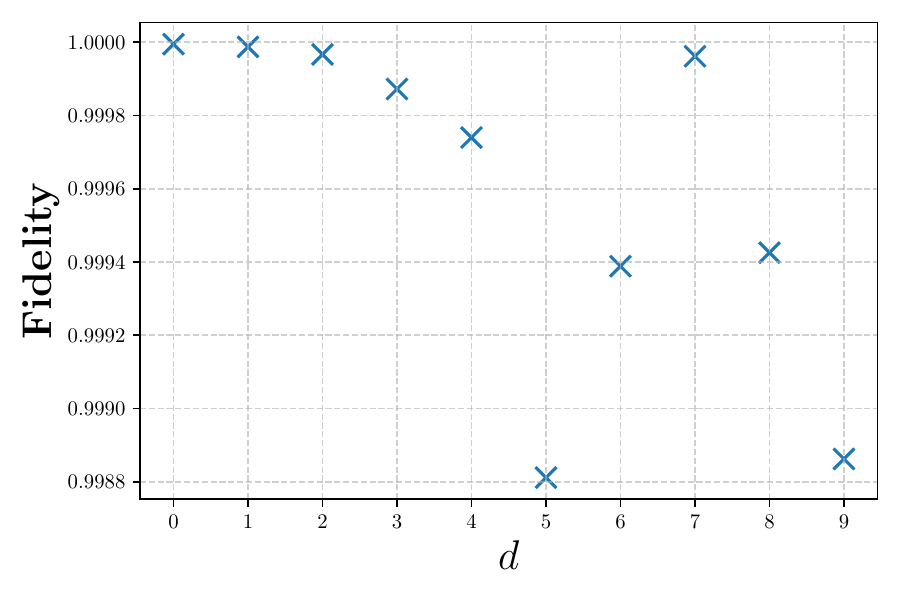}
    \caption{Fidelities between the the final states of the VQE and the exact ground states of the Hamiltonian for $J=0.5$, $m=2$ and $\varepsilon=0.5$ for different distance of the static charges.}
    \label{fig:static_string_breaking_fidelities_1}
\end{figure}

Using these initial states for simulating the VQE classically, we obtain the results shown in Fig.~\ref{fig:static_string_breaking_1} in the main text. For the case of the classical simulation using the state vector, we can also evaluate the fidelity of the final state of the VQE with the exact solution. These are shown in Fig.~\ref{fig:static_string_breaking_fidelities_1}.
As the results demonstrate, we are able to obtain fidelities above $99.88\%$ for all distances we study.

\section{Zero noise extrapolation\label{section:ZNE}}
ZNE is an error mitigation technique used to estimate the ideal values of observables~\cite{GiurgicaTiron2020,Ritajit2023}. The underlying idea is to measure the same observable at different noise levels, in order to perform an extrapolation to the zero noise-limit. In our experiments we use the ``gate folding'' method for the noise amplification. To this end, a unitary gate $U$ is replaced by $UU^{\dagger}U$. In an ideal, noise-free setting this will keep the result unchanged, however, in the presence of noise where each gate is realized imperfectly, this will increases the noise level. By systematically repeating (or ``folding'') certain gates to amplify noise, one can measure the expected value of an observable at different noise levels $\lambda$. 

\begin{figure}[htp!]
    \centering
    \includegraphics[width=\linewidth] {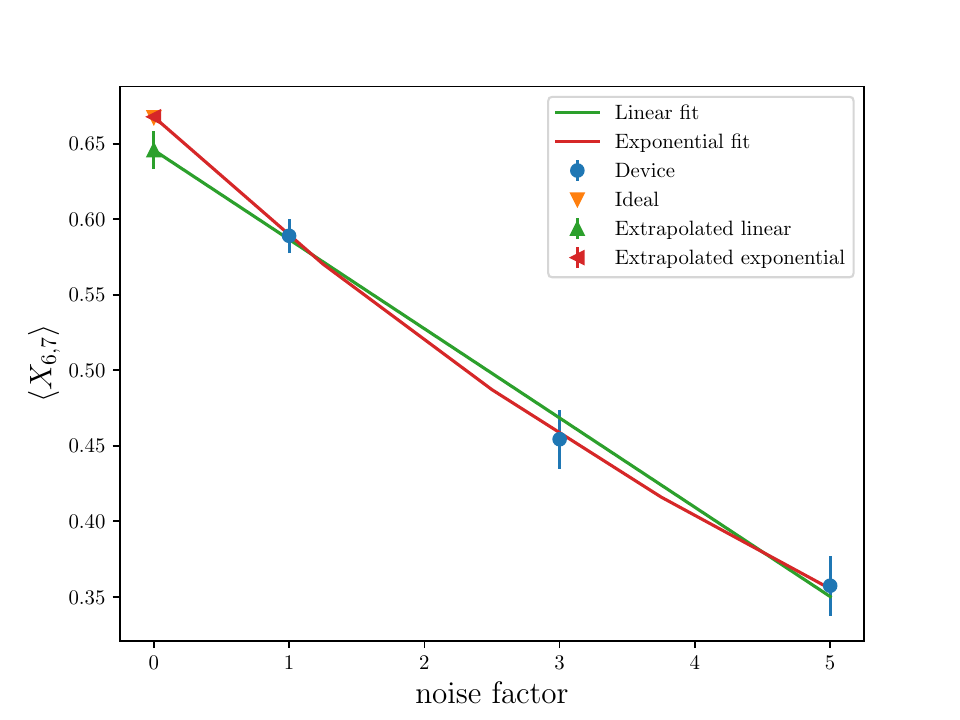}    
    \caption{Example of the ZNE for data obtained on \texttt{ibm\_marrakesh} measuring the expected value of the Pauli-$X$ operator acting on the 7th link of the chain for $J=0.5$, $m=0.1$ and $\varepsilon=0.15$. The blue points represent expectation values of the observable under increasing noise amplification factors 1, 3, and 5.
    The orange point represents the exact value of the Pauli operator at that site derived with an ideal simulation.
    The green and red curves are a linear and exponential fit to the noisy data points with and the green and red points being the extrapolated values  at the zero noise factor limit respectively.}
    \label{ZNE_example}
\end{figure}

A priori, the form of the fit function as well as the optimal noise amplification factors are not known beforehand. Here we follow standard practices and try linear as well as exponential fits~\cite{GiurgicaTiron2020,Ritajit2023}. The latter is motivated by the fact that under a global depolarizing channel an exponential decay with noise strength is expected, whereas the former should provide a good approximation for small values of noise. For the results of this paper we use the noise amplification factors of 1, 3 and 5 corresponding to the number of gates each unitary is replaced with (i.e.\ a factor of 1 corresponds to $U$, 3 to $UU^\dagger U$, etc.). Figure~\ref{ZNE_example} shows an example the expectation value of single Pauli operator.
For this example we see that our data follows a linear trend, however, an exponential fit is also capable of capturing the behavior well. 

Furthermore, we also explored probabilistic error amplification (PEA)~\cite{Kim2023a} for creating different noise-levels, which, compared to gate folding, should provide an unbiased estimator. However, for the regimes we study, we found that PEA provided less consistent results then gate folding, despite requiring a lot more computational time on quantum hardware.

\section{Reducing the parameters for larger scale simulations\label{sec:symmetry}}

For the system sizes explored in the main text, the number of parameters in the Ansatz circuit remains small enough to allow for direct optimization. However, as system sizes increase, the number of parameters in the Ansatz eventually grows to a point where the optimization procedure becomes challenging. While variants of qubit-ADAPT VQE have been proposed for constructing an Ansatz whose parameters can be systematically extrapolated to larger systems~\cite{Farrell2023,Gustafson2024}, in this section, we explore an alternative approach based on simple symmetry arguments to reduce the number of independent parameters. Specifically, we analyze modifications to the Ansatz circuit that enable efficient simulations of larger lattices, and, as we demonstrate below, allow for keeping a good level of accuracy.

\subsection{Translation invariance}
\begin{figure}[htp!]
    \centering
    \includegraphics[width=0.2\textwidth]{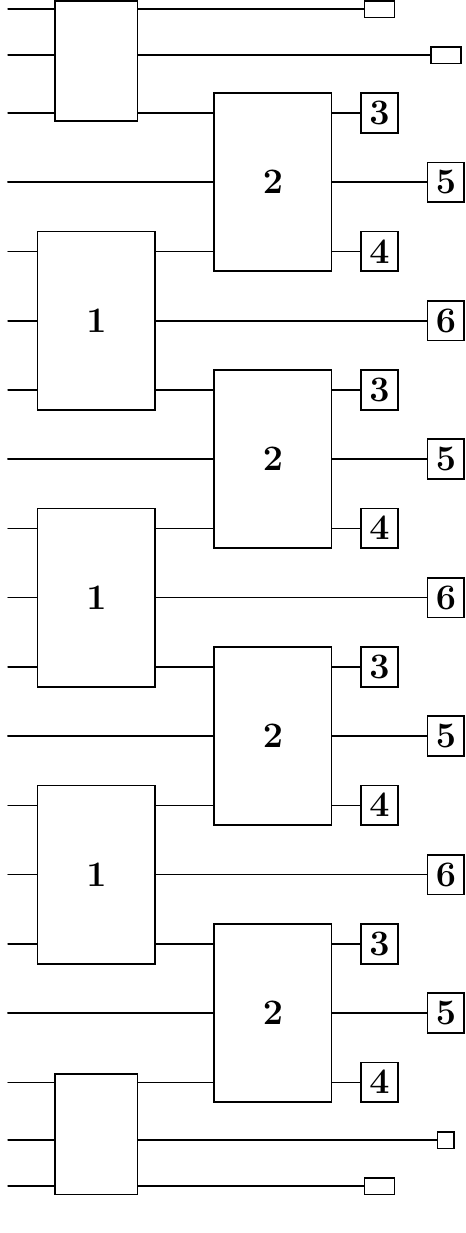}
    \caption{One layer of the Ansatz circuit with restricted parameters to enforce translational symmetry in the bulk region and to reduce the number parameters. Gates with identical numbers share the same variational parameter, reducing the overall parameter count compared to the Ansatz in presented in the main text. Due to the staggered formulation, translational symmetry in the bulk implies that a translation of two matter sites (or equivalently four qubits) leaves the bulk structure invariant. The parameters at the boundaries are not restricted to take into account boundary effects, as we work with open boundary conditions.
    }
    \label{fig:circuit_symmetry}
\end{figure}

In the absence of static charges, the bulk region of the lattice is expected to exhibit translational symmetry, where we have to translate the system by two matter sites due to the staggered formulation. Leveraging this property, we impose constraints on the individual parameters to enforce this symmetry. While we keep the parameters at the boundaries of the circuit unrestricted to account for boundary effects, we set the corresponding gate parameters for odd (even) sites as well as for the the links following odd (even) sites to be identical. This approach effectively reduces the number of independent variational parameters, thus simplifying the optimization landscape for the classical optimizer. Figure~\ref{fig:circuit_symmetry} presents a schematic representation of our Ansatz circuit, consisting of three-qubit gates corresponding to the kinetic term, and single-qubit gates for the mass and electric field terms. In the figure, we highlight the gates that share the same parameters with an the same number.
This symmetry-based parameter reduction approach provides a systematic and scalable way to construct variational circuits for larger lattices, ensuring computational feasibility while preserving the essential structure of the ground state.

\subsection{Removing redundant parameters by separating layers \label{app:symmetries}}

\begin{figure*}[htp!]
    \centering
    \includegraphics[width=0.98\textwidth]{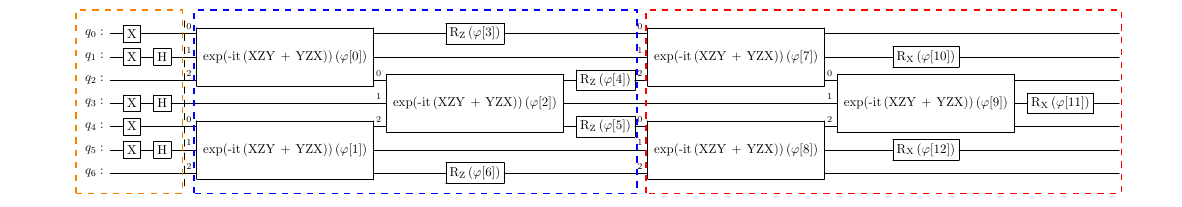}
\caption{Ansatz circuit where the single-qubit gates acting on matter and electric field related gates are separated to different layers. Gates in the orange box are used to set $\ket{\psi_\text{init}}$ to the noninteracting ground state. The gates inside the blue and red boxes represent two different types of layers, which are alternated in the Ansatz.}
\label{fig:circuit_alt_no1stlayer}
\end{figure*}

A second approach to reducing the number of variational parameters in the circuit is to introduce a layered Ansatz, where the single-qubit gates acting on matter sites and link variables are placed in separate layers. This structuring allows for reduced parameters when considering multiple layers, while still preserving gauge invariance and, as we show in the next paragraph, enough flexibility to reliably capture the ground state of the model. Figure~\ref{fig:circuit_alt_no1stlayer} presents an example of this layered structure. Compared to the original Ansatz proposed in the main text (see Fig.~\ref{fig:circuit_no1stlayer} in the main text), the single-qubit gates acting on the matter and gauge links are distributed to alternating layers, hence we will refer to this Ansatz as the ``alternating Ansatz''.
By organizing the circuit in this way, we preserve gauge invariance while simultaneously reducing the number of independent parameters.

\subsection{Benchmarking the Ansatz circuits using translational symmetry and alternating layered structure}

To assess the impact of the symmetry-based parameter reduction and the layered Ansatz, we test the original circuit and the alternating structure, both with and without the symmetry constraints for the bulk region. For benchmarking, we focus on the relative error of the energy obtained from the VQE for a lattice with 10 matter sites, corresponding to 19 qubits, and Hamiltonian parameters $J=0.5$, $m=2$, and $\varepsilon=0.5$ in absence of static charges.

\begin{figure}[htp!]
    \centering
    \includegraphics[width=\linewidth]{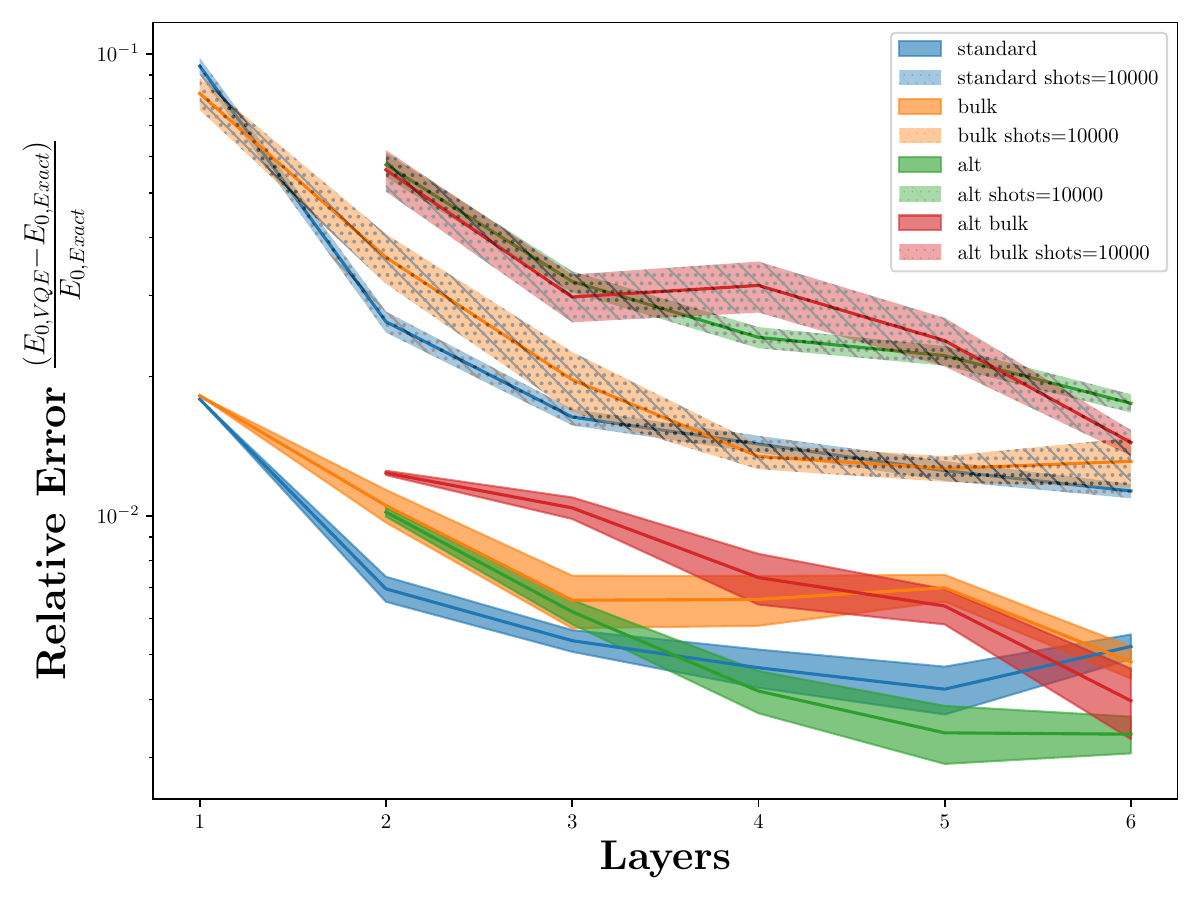}
    \caption{Average relative error of the ground state energy obtained from VQE as a function of the number of Ansatz layers for $L=10$, $J=0.5$, $m=2$, and $\varepsilon=0.5$. The solid lines surrounded by a homogeneous filled region represent the average results as well as the standard deviation from 90 a classical simulations using the state vector, the lines surrounded by a hatched area represent the data for classically simulating the VQE using 10,000 measurements per iteration. Different colors represent different ansätze, where blue is the original Ansatz from the main text, orange the Ansatz restricting the parameters in the bulk to incorporate translation invariance, green the Ansatz separating the single-qubit gates to different layers and red the Ansatz combining both parameter reduction techniques.}
    \label{fig:error_vs_layers}
\end{figure}

\begin{figure}[htp!]
    \centering
    \includegraphics[width=\linewidth]{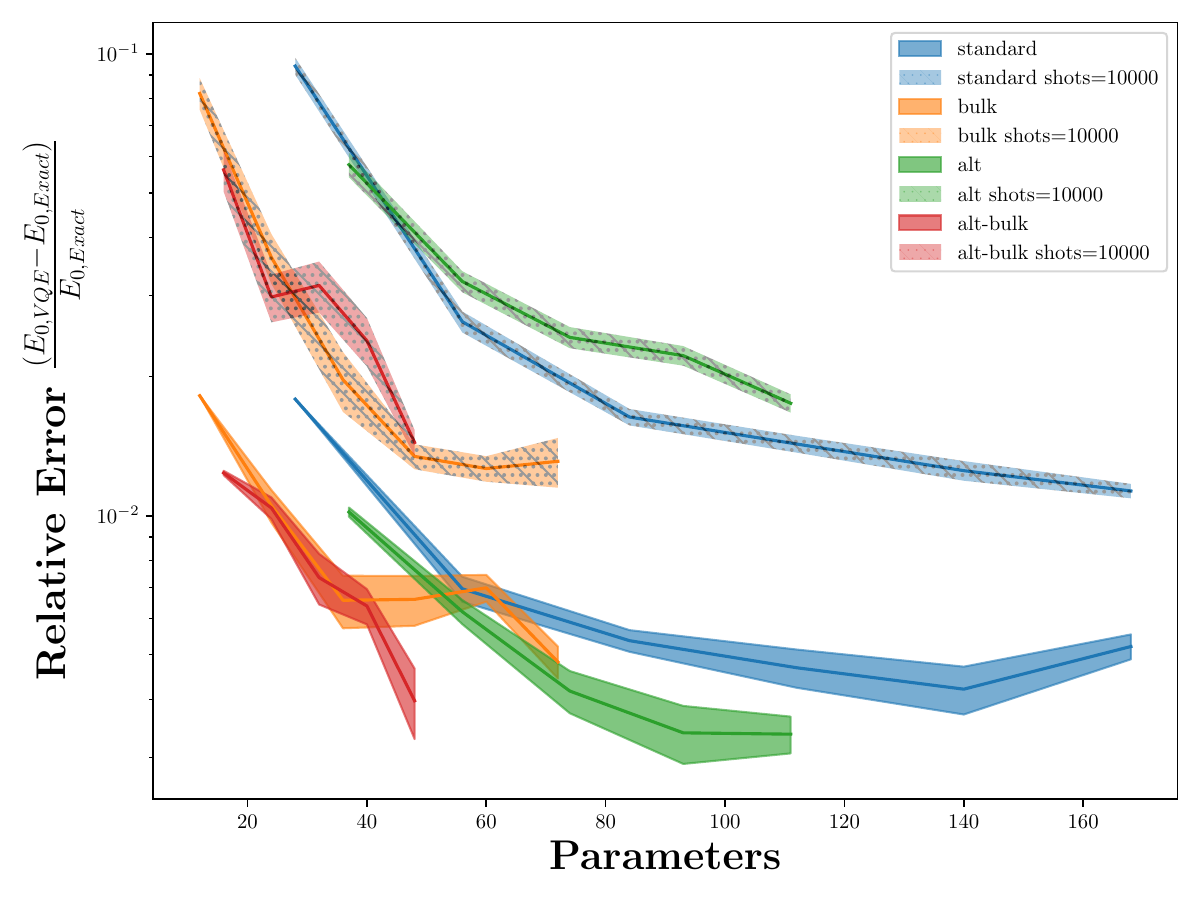}
    \caption{Same data as in Fig.~\ref{fig:error_vs_parameters} plotted as a function of the corresponding number of parameters.}
    \label{fig:error_vs_parameters}
\end{figure}

In Figs.~\ref{fig:error_vs_layers} and \ref{fig:error_vs_parameters}, we present our results for the relative error of the energy plotted as both a function of the number of layers and parameters respectively for better comparability of the different types of ansätze. For all types of ansätze we used the same small-angle initialization as in the main text and performed 100 independent runs. From these results we exclude the lowest-performing 10 runs to remove cases where the optimization gets directly trapped in local minima. The solid lines represent the average out of the 90 remaining runs, and the shaded regions indicate the standard deviation.

Focusing on Fig.~\ref{fig:error_vs_layers} first, we see that the relative error in the energy is generally decreasing with an increasing number of layers. For the classically simulated VQE using the state vector, the result for all types of ansätze are of the same order of magnitude, with the Ansatz presented in the main text and alternating one without putting any constraints on the parameters doing slightly better then the ones enforcing translation invariance in the bulk. In the presence of shot noise, we observe that the original Ansatz performs best, where differences between keeping all parameters independent and imposing constraints due to translation invariance are small. 

Looking at the same results, but as a function of number of parameters in the Ansatz (c.f.\ Fig.~\ref{fig:error_vs_parameters}), we see a considerable difference in the performance with the number of parameters. While as a function of layers one of the symmetry reduced ansätze performed similar or even better than the original one for both, using the state vector and simulating a finite number of measurements, the parameter reduced ansätze generally provide better performance with a smaller number of parameters. 

These results demonstrate that leveraging symmetry arguments and alternating layers provides an efficient way to reduce the number of parameters while maintaining an accurate representation of the ground state. In particular, such a parameter reduction provides an important avenue towards scaling the VQE up to larger system sizes.

\FloatBarrier
\bibliography{papers}

\end{document}